\documentclass[sigconf]{acmart}

\AtBeginDocument{%
  \providecommand\BibTeX{{%
    \normalfont B\kern-0.5em{\scshape i\kern-0.25em b}\kern-0.8em\TeX}}}

\setcopyright{none}
\copyrightyear{2018}
\acmYear{2018}
\acmDOI{10.1145/1122445.1122456}

\acmConference[SC '21]{Supercomputing '21: The International Conference for High Performance Computing, Networking, Storage, and Analysis}{Nov 14–19, 2021}{St. Louis, MO}
\acmBooktitle{Supercomputing '21: The International Conference for High Performance Computing, Networking, Storage, and Analysis, Nov 14–19, 2021, St. Louis, MO}
\acmPrice{15.00}
\acmISBN{XXX-X-XXXX-XXXX-X/XX/XX}

\usepackage{soul} 
\usepackage{hhline}
\usepackage{caption}
\usepackage{subcaption}
\usepackage{amsmath}
\usepackage{bm}
\usepackage{xspace}

\newcommand{\dani}[1]{}

\usepackage{array}
\newcolumntype{P}[1]{>{\centering\arraybackslash}p{#1}}

\newcommand{\flare}{\textsc{Flare}\xspace}

\newcommand{\up}{\includegraphics[scale=0.4,trim=0 0 0 0]{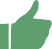}}
\newcommand{\down}{\includegraphics[scale=0.4,trim=0 2 0 0]{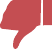}}
\newcommand{\half}{\includegraphics[scale=0.4,trim=0 2 0 0]{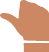}}

\newcommand{\uptext}{\includegraphics[scale=0.5,trim=0 0 0 0]{figs/up.pdf}}
\newcommand{\downtext}{\includegraphics[scale=0.5,trim=0 2 0 0]{figs/down.pdf}}
\newcommand{\halftext}{\includegraphics[scale=0.5,trim=0 2 0 0]{figs/half.pdf}}

\graphicspath{ {./figs/}}

\definecolor{r1}{RGB}{87,114,158}
\definecolor{r2}{RGB}{204,137,99}
\definecolor{r3}{RGB}{93,157,107}
\definecolor{r4}{RGB}{196,78,82}
\definecolor{r5}{RGB}{129,114,180}
\usepackage{marginnote}
\usepackage{soul} 
\usepackage{hhline}
\definecolor{lightyellow}{RGB}{250, 250, 180}
\definecolor{HLYELLOW}{RGB}{240, 127, 0}
\definecolor{hlyellow}{RGB}{240, 127, 0}
\sethlcolor{lightyellow}
\definecolor{lightcyan}{RGB}{160,255,255}
\usepackage{xparse}

\usepackage{mdframed}
\global\mdfdefinestyle{review}{%
linecolor=lightyellow,linewidth=3pt,%
leftmargin=0cm,rightmargin=0cm,%
skipabove=0cm,skipbelow=0cm,%
innerrightmargin=0cm,innerleftmargin=0cm,%
innerbottommargin=0cm,innertopmargin=0cm,%
backgroundcolor=lightyellow
}
\global\mdfdefinestyle{reviewtext}{%
linecolor=lightyellow,linewidth=0pt,%
leftmargin=0cm,rightmargin=0cm,%
skipabove=0.1cm,skipbelow=0.1cm,%
innerrightmargin=0cm,innerleftmargin=0cm,%
innerbottommargin=0cm,innertopmargin=0cm,%
backgroundcolor=lightyellow
}

\begin{document}

\DeclareDocumentCommand\review{m g g}{%
    {\IfNoValueF {#2}{%
    \IfNoValueF {#3}{%
    {\marginnote{\sethlcolor{#3}\hl{\normalfont \textbf{{\normalsize{\color{white}#2}}}}}%
    }%
    }%
    \IfNoValueT {#3}{%
    {\marginnote{\normalfont \textbf{\normalsize{#2}}}%
    }%
    }%
    }%
    \hl{#1}%
    }%
}

\title{Flare: Flexible In-Network Allreduce}

\author{Daniele De Sensi}
\email{daniele.desensi@inf.ethz.ch}
\affiliation{
  \institution{ETH Zurich}
  \city{Zurich}
  \country{Switzerland}
}

\author{Salvatore Di Girolamo}
\email{salvatore.digirolamo@inf.ethz.ch}
\affiliation{
  \institution{ETH Zurich}
  \city{Zurich}
  \country{Switzerland}
}

\author{Saleh Ashkboos}
\email{saleh.ashkboos@inf.ethz.ch}
\affiliation{
  \institution{ETH Zurich}
  \city{Zurich}
  \country{Switzerland}
}

\author{Shigang Li}
\email{shigang.li@inf.ethz.ch}
\affiliation{
  \institution{ETH Zurich}
  \city{Zurich}
  \country{Switzerland}
}

\author{Torsten Hoefler}
\email{torsten.hoefler@inf.ethz.ch}
\affiliation{
  \institution{ETH Zurich}
  \city{Zurich}
  \country{Switzerland}
}

\renewcommand{\shortauthors}{De Sensi et al.}

\begin{abstract}
  The \textit{allreduce} operation is one of the most commonly used communication routines in distributed applications. To improve its bandwidth and to reduce network traffic, this operation can be accelerated by offloading it to network switches, that aggregate the data received from the hosts, and send them back the aggregated result. However, existing solutions provide limited customization opportunities and might provide suboptimal performance when dealing with custom operators and data types, with sparse data, or when reproducibility of the aggregation is a concern. To deal with these problems, in this work we design a flexible programmable switch by using as a building block PsPIN, a RISC-V architecture implementing the sPIN programming model. We then design, model, and analyze different algorithms for executing the aggregation on this architecture, showing performance improvements compared to state-of-the-art approaches.
\end{abstract}




\maketitle

\section{Introduction}
\textit{Allreduce} is a commonly used collective operation where $P$ vectors, one for each host participating in the operation, are aggregated together. If each vector contains $Z$ elements, the allreduce operation aggregates the $P$ vectors element-wise and returns to each host a vector of $Z$ aggregated elements. Common aggregation functions include the sum of the elements, the computation of their minimum or maximum, and others~\cite{mpi}. Allreduce is widely used in many applications, including scientific applications~\cite{mpi-characterization,milc}, deep learning~\cite{10.1145/3320060}, graph processing~\cite{6957236,DBLP:journals/corr/ZhaoC13}, big data analytics~\cite{10.1007/978-3-642-03770-2_30}, and others, and recent studies~\cite{mpi-characterization} show that \textit{``\texttt{MPI\_Allreduce} is the most significant collective in terms of usage and time''}.

The simplest bandwidth-optimal allreduce algorithm is the Rabenseifner algorithm (also known as \textit{ring allreduce})~\cite{bwoptimalallreduce}. This algorithm is composed of two phases: a \textit{scatter-reduce}, and an \textit{allgather} phase. $P$ hosts are arranged into a logical ring, and in each of these two phases, each host sends to its neighbor $(P - 1)$ messages, each of size $\frac{Z}{P}$ (where $Z$ is the number elements to be reduced). The total amount of data sent by each host is then $2(P - 1)\frac{Z}{P} \approx 2Z$. To reduce the amount of transmitted data, and thus increase the performance, hosts can exploit \textit{in-network} compute, i.e., they can offload the allreduce operation to the switches in the network. 

To outline the advantages of performing an in-network allreduce, we describe the general idea underlying most existing in-network reduction approaches~\cite{sharp,nvidiaar,switchml}. We first suppose to have the $P$ hosts connected through a single switch. Each of the hosts sends its data to the switch, that aggregates together the vectors coming from all the hosts, and then sends them back the aggregated vector. Differently from the host-based optimal allreduce, in the in-network allreduce each host only sends $Z$ elements, thus leading to a 2x reduction in the amount of transmitted data. If the switches can aggregate the received data at line rate, this leads to a \textbf{2x bandwidth improvement} compared to a host-based allreduce. Besides improvements in the bandwidth, in-network allreduce also reduces the network traffic. Because the interconnection network consumes a large fraction of the overall system power (from 15\% to 50\% depending on the system load~\cite{10.1145/1816038.1816004}), any reduction in the network traffic would also help in reducing the power consumption and thus the running cost of the system. 

\begin{figure}[htb]
    \centering
    \includegraphics[width=\columnwidth]{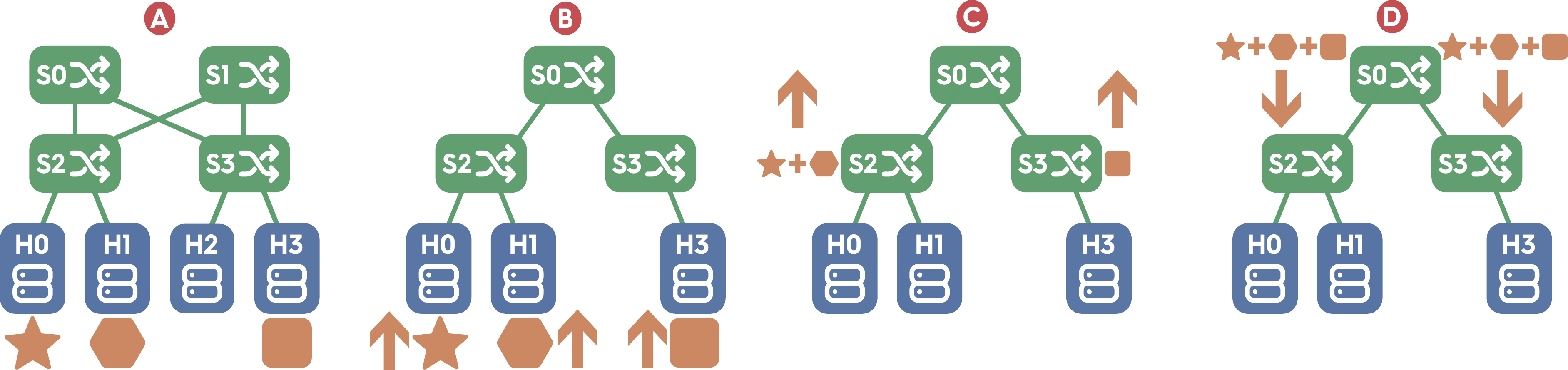}
    \caption{Example of an in-network allreduce.}
    \label{fig:intro}
\end{figure}

If the hosts participating in the reduction span across multiple switches, the aggregation can be done recursively, as shown in Figure~{\ref{fig:intro}}. Let us suppose the hosts \textit{H0}, \textit{H1}, and \textit{H3} need to perform an allreduce on their data, denoted through geometric figures ({\includegraphics[scale=0.04,trim=0 30 0 0]{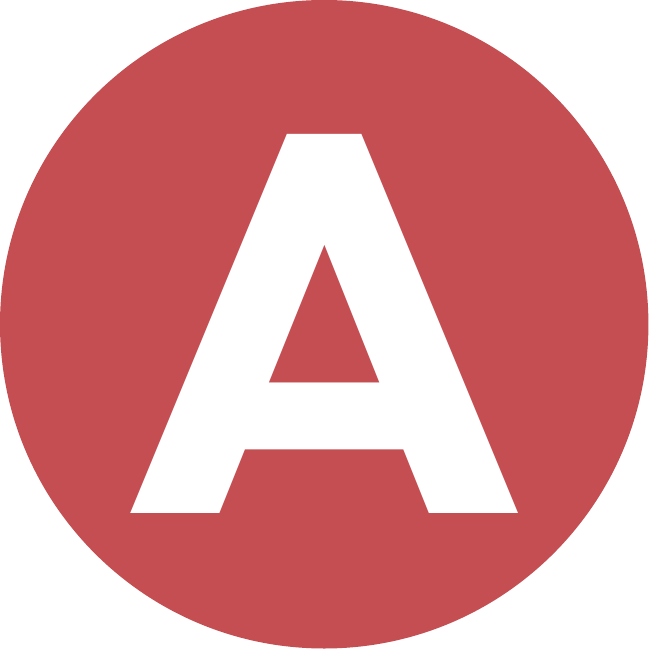}}). First, they build a \textit{reduction tree}, where the leaves are the hosts and the intermediate nodes are a subset of the switches (\textit{S0}, \textit{S2}, and \textit{S3} -- {\includegraphics[scale=0.04,trim=0 30 0 0]{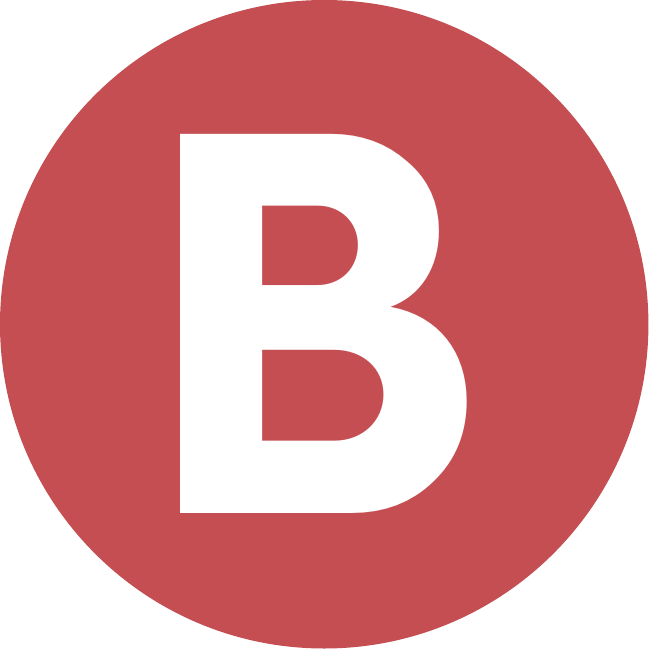}}). After a switch aggregates all the data coming from its children, if it is an intermediate switch in the tree, it sends the aggregated data to its parent ({\includegraphics[scale=0.04,trim=0 30 0 0]{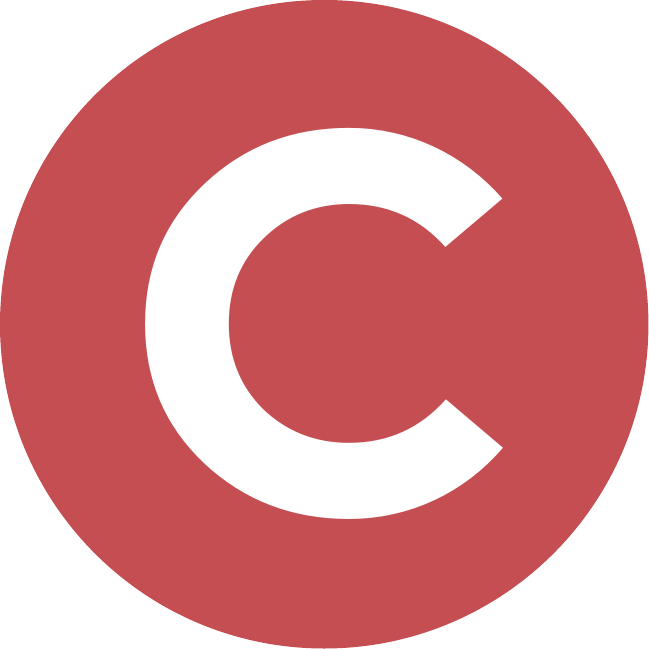}}). Otherwise, if it is the root of the tree, it broadcasts the data down the tree, and the fully reduced data will eventually reach all the hosts ({\includegraphics[scale=0.04,trim=0 30 0 0]{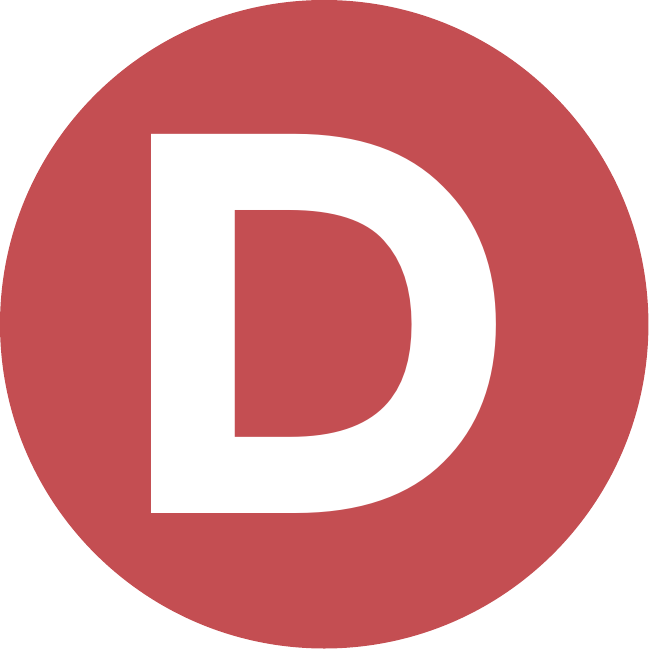}}).



Although many in-network allreduce solutions have been designed in the years, especially for HPC networks, all of them lack flexibility, in terms of supported data types, data formats, and operators, as we show in Section~{\ref{sec:state}}. We show in this work that this limits the applicability of in-network reduction (for example, none of them can deal with sparse data) and can lead to performance degradation. To overcome these limitations, we propose \flare (\textit{\underline{Fl}exible In-Network \underline{A}ll\underline{re}duce}). \flare includes a programmable switch based on PsPIN~\cite{pspin}, and a set of aggregation algorithms for exploiting the switch architecture at best. PsPIN is an open-source multi-cluster RISC-V architecture~\cite{Waterman:EECS-2014-54}, implementing the sPIN programming model~\cite{spin}, and allowing the programmer to specify \textit{packet handlers} (i.e., the code to be executed for each packet), as plain C functions. The packet handlers can be programmed and loaded by the system administrator after the switch has been deployed. This extends the range of applicability and eases the programmer's task compared to existing programmable switches. 

In this work, we answer the following research questions:
\begin{description}
\item[Q1] What are the limitations of existing in-network data reduction solutions and why there is the need for flexible in-network reductions? (Section~\ref{sec:state})
\item[Q2] How to design a switch architecture for flexible in-network data reduction? (Section~\ref{sec:switch})
\item[Q3] How to exploit such architecture to efficiently execute allreduce operations? (Sections~\ref{sec:design}-\ref{sec:sparsity})
\item[Q4] What performance advantages can we expect compared to existing in-network and host-based solutions? (Sections~\ref{sec:evaluation:types} and~\ref{sec:evaluation:sparse})
\end{description}

\section{State of the art}\label{sec:state}
Existing solutions for in-network reductions can be divided into three broad categories: those implemented on fixed-function (i.e., non programmable) switches, those relying on \textit{Field-Programmable Gate Arrays} (FPGAs), and those targeting programmable switches.
\subsection{Fixed-function switches}
Solutions targeting fixed-function switches~\cite{sharp,sharp2,crayxcpdf,tofu,percs,percspatent,anton,nvidiaar} are characterized by high performance but tied to a specific implementation, and do not provide any customization opportunity. Many HPC networks provide some kind of support for in-network allreduce. For example, SHARP~\cite{sharp,sharp2} can be used on Mellanox's networks, and similar solutions are provided on Tofu~\cite{tofu}, and Cray's Aries networks~\cite{crayxcpdf}. Usually, these solutions support the most commonly used MPI reduction operators, on both integer and floating point data, but some of them might only support latency-sensitive small data reductions~\cite{tofu,crayxcpdf}. 

\subsection{FPGAs}
Another set of solutions targets \textit{Field-Programmable Gate Arrays} (FPGAs)~\cite{panama,netreduce}, trading performance for some flexibility. Some of these solutions~\cite{panama} work by placing the FPGA between the hosts and the switch. Other solutions instead connect the FPGA only to the switch~\cite{netreduce}, and configure the switch to route the packets that need to be aggregated to the FPGA. After the FPGA aggregates the packet, it sends the result back to the switch, that will then send it to the next hop. However, even if more flexible than fixed-function switches, extending an FPGA-based in-network allreduce solution might still require a hardware re-design. Moreover, the FPGAs considered by these solutions have limited network bandwidth, ranging from four 10Gbps ports~\cite{panama} to six 100Gbps ports~\cite{netreduce}.

\subsection{Programmable switches}
Last, some solutions target programmable switches~\cite{atp,switchml,omnireduce}, that provide more flexibility by allowing the programmer to specify the type of processing to be executed on the packets with high-level programming languages. Programmable switches are often implemented through \textit{Reconfigurable Match-Action Tables} (RMTs)~\cite{rmt,tofino,intelfm,8850761}, and can be configured with the P4 programming language~\cite{p4, survey-p4-1, survey-p4-2}. In such architecture, each packet traverses a pipeline of $10-20$ stages~\cite{10.1145/3317550.3321439}, each applying a longest-prefix match of one or more packet fields against a set of rules, to select an action to be executed on that field. Possible actions include simple \textit{Arithmetic Logic Unit} (ALU) operations, memory read/write, and modifications of packet fields and the packet destination. Both the matching rules and the actions to be executed can be customized by the programmer.

However, the existing programmable switches architecture is rigid and has several limitations~\cite{201474,10.1145/3317550.3321439}, including the lack of loops, and a limited set of operations. Operations with dependencies must be placed on subsequent pipeline stages and non-trivial applications cannot be mapped on such an architecture. Moreover, state management is limited, and everything needs to be expressed as a statically-sized array. Existing hardware can also only perform 32 operations per packet~\cite{switchml,atp}, and each packet can only access each memory location once. To overcome these limitations, packets can be \textit{``recirculated''}, i.e., sent back to the switch on a loopback port. This however reduces the bandwidth of the switch proportionally to the number of times each packet is recirculated. For example, to process the data sent by the hosts at 100Gbps, existing allreduce implementations for programmable switches only allow 16 ports to be used on a 64-port switch~\cite{switchml}. 

\subsection{Limitations}
We identified the following \textit{flexibility limitations} in existing in-network allreduce solutions, that we summarize in Table~\ref{tab:comparison}, and that we overcome in this work:

\paragraph{\textbf{F1 - Custom operators and data types}}
Solutions for fixed-function switches and FPGAs support a predefined set of operators and data types~\cite{sharp}, that cannot be customized nor extended (Table~{\ref{tab:comparison}}, \downtext). In-network aggregation solutions for programmable switches can be customized, but they are still limited by the hardware (\halftext). For example, existing programmable switches do not have floating-point units, and do not support multiplication/division for integer data~\cite{tofino,switchml,atp}. Moreover, due to the limited number of elements that can be processed per packet (independently from the element size), aggregating sub-byte elements, as common in deep learning applications~\cite{hoefler2021sparsity,quantization,10.5555/3018874.3018875,37631} would not improve application performance. On the other hand, \flare provides full customizability of operators and data types (\uptext), allowing the user to specify arbitrary aggregation functions as sPIN handlers (Section~\ref{sec:switch}). 

\paragraph{\textbf{F2 - Sparse data}} Many applications need to reduce sparse data~\cite{hoefler2021sparsity,sparcml,omnireduce,6957236}, i.e., data containing mostly null values. To save bandwidth and improve performance, an application might only transmit and reduce the non-null values. However, to the best of our knowledge, none of the solutions targeting fixed-function switches provide explicit support for sparse data (\downtext). Among the programmable switches solutions, only one partially targets in-network sparse data reduction~\cite{omnireduce}. However, it forces the application to sparsify the data per-block, i.e., to send sparse blocks of dense data (\halftext). This is however not possible for any application and, even when possible (for example for deep learning models training), this negatively affects the convergence of the training~\cite{omnireduce}. On the other hand, with \flare we design the first in-network sparse allreduce algorithm, that does not make any assumption on the data sparsity, and can process the data as generated by the hosts (\uptext), improving application performance compared to existing solutions (Section~\ref{sec:evaluation:sparse}). 

\paragraph{\textbf{F3 - Reproducibility}} Many scientific applications require the computed results to be reproducible across different runs and for different allocations. For example, in weather and climate modeling, a small difference in computation on the level of a rounding error could lead to a completely different weather pattern evolution~\cite{bams-htor,weather}. However, some aggregation functions (e.g., floating-point summation~\cite{Geyer2021,COLLANGE201583,6831947,6812157,Villa_effectsof}) might depend on the order in which the elements are aggregated, and the final result might change if, in subsequent runs, packets arrive at the switch in a different order. Many solutions for fixed-function switches ensure reproducibility, in most cases by storing all the packets and aggregate them in a pre-defined order only when they are all present~\cite{sharp,panama}. This however increases the memory occupancy of the switch, even when this is not required by the application, for example for integer data or for an application that might tolerate different results across different runs (\uptext). 
Differently from existing solutions, \flare guarantees reproducibility only when explicitly requested by the user, by organizing the aggregation in the switch so that associativity of the operator is never used. Moreover, \flare reproducible allreduce does not require storing all the packets before aggregating them (\uptext). 

 \begin{table}[htpb]
 \centering
\footnotesize
\resizebox{\columnwidth}{!}{%
 \begin{tabular}{|c||c|c|c|c|c|c|c||c|c||c|c|c|c|} 
 \hline
              & \multicolumn{7}{c||}{\textsc{Fixed-Function Switches}} & \multicolumn{2}{c||}{\textsc{FPGAs}} & \multicolumn{4}{c|}{\textsc{Progr. Switches}} \\ \hline
              & \cite{sharp} & \cite{sharp2} & \cite{crayxcpdf} & \cite{tofu} & \cite{percs}             & \cite{anton} & \cite{nvidiaar} & \cite{panama} & \cite{netreduce} & \cite{atp} & \cite{switchml} & \cite{omnireduce} & \textsc{\textbf{Flare}} \\ \hline\hline
  \textsc{\textbf{F1}} &  \down       & \down         & \down            & \down       & \down                    & \down        & \down           & \down         & \down            & \half      & \half           & \half             & \up                   \\ \hline
  \textsc{\textbf{F2}} &  \down       & \down         & \down            & \down       & \down                    & \down        & \down           & \down         & \down            & \down      & \down           & \half             & \up                   \\ \hline
  \textsc{\textbf{F3}} &  ?           & \up         & ?                & ?           & \up                    & ?            & \down           & \up         & \up            & \down      & \down           & \down             & \up                   \\ \hline
 \end{tabular}
 }
 \caption{Comparison between existing in-network allreduce solutions. F1: Custom operators and data types, F2: Sparse data, F3: Reproducibility. \uptext : provided, \halftext: partially provided, \downtext: not provided, ?: unknown. \dani{improve columns alignment}}
 \label{tab:comparison}
\end{table}

\section{Switch Architecture}\label{sec:switch}
\begin{figure}[ht]
    \centering
    \includegraphics[width=\columnwidth]{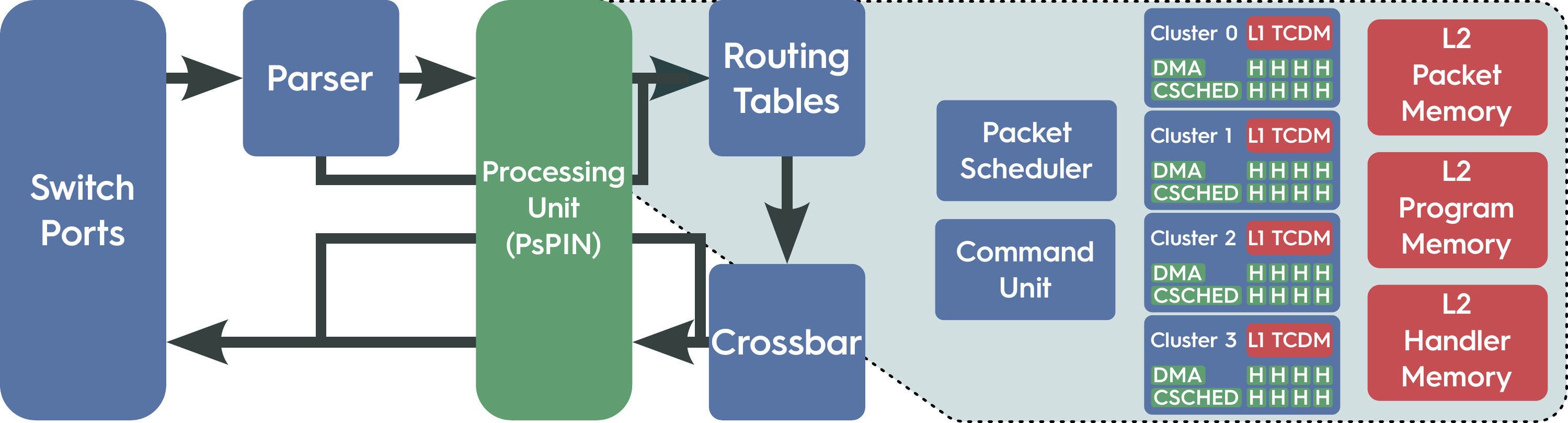}
    \caption{PsPIN switch high-level architecture.}
    \label{fig:switch}
\end{figure}

Figure~\ref{fig:switch} illustrates the high-level architecture of a \flare PsPIN-based programmable switch. After a packet is received from any of the switch ports, its headers are processed by a \textit{parser}~\cite{ZOLFAGHARI2020102910,6665172} that, based on configurable \textit{matching rules}, decides if the packet must be processed by a \textit{processing unit} (or sent directly to the routing tables unit\footnote{By doing so, packets that do not need additional processing are not further delayed.}), and which function must be executed on the packet. The processing unit can modify the content of each packet, and decide its destination. We assume that the network administrator configures the matching rules in the parser through the control plane~\cite{sdn}, specifying the functions to execute for each packet, based on the values of specific packet fields (e.g., \textit{EtherType} in the Ethernet header~\cite{rfc894}, or IP optional headers). In principle, the code to be executed could be specified by users of the system, but this would open security and accountability concerns that are outside the scope of this paper.

Differently from existing programmable switches~\cite{p4,tofino,intelfm,domino,flowblaze}, that implement the processing unit through \textit{Reconfigurable Match Action} (RMT) tables~\cite{rmt,8850761}, we consider the processing unit to be implemented as a PsPIN unit~\cite{pspin}, highlighted in the right part of Figure~\ref{fig:switch}. PsPIN is a clustered RISC-V built on top of the PULP~\cite{pulp} platform. If a packet needs to be processed, the parser copies the packet in a 4MiB \textit{L2 packet memory} and sends a request to a packet scheduler\footnote{If the packet memory is full, the packet is dropped or congestion is notified before filling the buffer, depending on the specific network where the switch is integrated.}. The packet scheduler forwards the request to one of multiple clusters (four in the example). A cluster-local scheduler (CSCHED) then selects a \textit{Handler Processing Units} (HPU, eight in the example, denoted with \includegraphics[scale=0.6,trim=0 0 0 0]{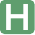}) where the packet will be processed, and starts a DMA copy of the packets from the L2 packet memory to a single-cycle scratchpad 1MiB memory (L1 TCDM - \textit{Tightly-Coupled Data Memory}). The cluster scheduler also loads the code to be executed from the 32KiB \textit{L2 program memory} to a cluster-local 4KiB instruction cache (not shown in the figure). Each HPU is a RISC-V RI5CY core~\cite{7864441} that can execute sPIN \textit{handlers}~\cite{spin} i.e., C functions defining how to process the content of the packet. The handlers can use the single-cycle L1 memory also to store and load data, that is preserved after the processing of the packet is terminated, and until the handler is uninstalled from the control plane. Each cluster also has a DMA engine that can be used to access a globally shared 4MiB memory (\textit{L2 handler memory}). 

After processing a packet, each HPU can set its destination and send it to the routing tables unit through a command unit. After getting the destination port, the packet is sent to a crossbar unit, that also implements queueing and quality of service (QoS) functionalities. This unit is also known as \textit{traffic manager}~\cite{tofino}, and we assume that the implementation and functionalities of both the routing tables unit and the crossbar are similar to those of any other switch. When the packet is ready to be sent, it might be processed again by the processing unit\footnote{For example, this might be useful when performing telemetry, to insert in the packet information about queue occupancy or queueing time.}. After this additional and optional processing step, the packet can be either dropped or forwarded.

The PsPIN processing unit is clocked at 1 GHz, and a PsPIN unit with four clusters with eight cores each (as the one shown in the example) occupies 18.5 mm$^2$~\cite{pspin} in a 22nm FDSOI process. Half of the area in the PsPIN unit is occupied by the L2 memory, and each cluster occupies on average 2mm$^2$~\cite{pspin}, most of which is L1 memory (1.64mm$^2$). We also add an FP32/FP16 \textit{Floating Point Unit} (FPU)~\cite{fpnew} to each core, increasing the area of the cluster to 2.29mm$^2$. The processing unit in programmable switches is estimated to occupy up to 140mm$^2$ in a 28nm FDSOI process~\cite{arealong}. By scaling the area, we set a 180mm$^2$ target area for our PsPIN unit, and we then assume we can fit $\sim$64 clusters and the L2 memory in our processing unit area budget. We assume that more clusters can be fit in the unit by organizing them hierarchically (in principle, by having separate units with a scheduler on the front). Moreover, it is worth remarking that our estimation of the available area budget is conservative because existing switches are manufactured in 16nm~\cite{slingshot} to 7nm~\cite{tomahawk4} processes. 

\section{Flare Allreduce General Design}\label{sec:design}
In \flare, before starting the aggregation, the application sends a request to a network manager node~\cite{sharp,sharp2}, that computes a reduction tree, and installs the handlers on all the switches of the tree through the control plane of the network. For each switch of the reduction tree, the network manager also sets the number of ports from which it will receive the packets to be aggregated (i.e., its children in the reduction tree), and the port to be used to forward the aggregated data (i.e., the port connected to the parent in the reduction tree). Because the structure of the reduction tree depends on the location of the hosts participating in the reduction, this setup phase must be done once for each subset of hosts executing an allreduce. For example, in the case of MPI, this must be done once for each communicator, similarly to other existing approaches~\cite{sharp,sharp2}. 

Each switch can participate simultaneously in different allreduces (issued by the same user/application or by different ones), and the network manager assigns a unique identifier to each allreduce, so that only packets belonging to the same allreduce are aggregated together. Because each allreduce consumes some memory resources on the switch, we assume that the memory is statically partitioned across a predefined maximum number of allreduces, as done by most in-network reduction approaches~\cite{sharp,sharp2,crayxcpdf,tofu,switchml,omnireduce}. If any switch on the reduction tree is already managing the maximum possible number of allreduces, the network manager can try to recompute a different reduction tree excluding that switch or just reject the request issued by the application, which then needs to rely on host-based allreduce~\cite{sharp,sharp2}. Because the memory is partitioned and the compute resources are dynamically assigned, to simplify the analysis in the following we focus on describing what happens for a single allreduce. 

\begin{figure}[htb]
    \centering
    \includegraphics[width=.7\columnwidth]{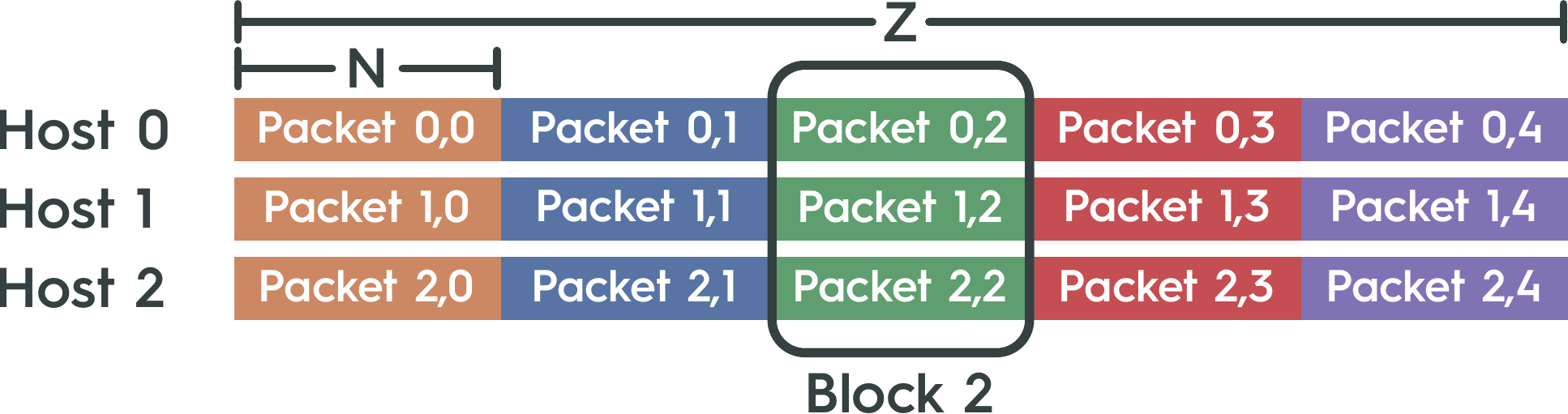}
    \caption{Partitioning of the data into packets and \textit{aggregation blocks}.}
    \label{fig:blocks}
\end{figure}

Once the setup is complete, the hosts can start sending the data to be reduced. We assume that $Z$ elements must be aggregated and that each host splits its data in $\frac{Z}{N}$ packets of $N$ elements each. In Figure~\ref{fig:blocks} we show an example where we have three hosts, and each of them splits the data into five packets. To simplify the exposition we denote each packet with two indexes: the index of the host that generated the packet, and the position of the packet within the host data. For example, \textit{Packet 0,1} is the second packet sent by \textit{Host 0}. The size of the packets is smaller than the \textit{Maximum Transmission Unit} (MTU) and, besides the payload, the hosts also add a small header containing the identifier of the allreduce (not shown) and of the packet within that allreduce. We refer to a set of packets to be aggregated together as a \textit{reduction block} (\textit{block} for brevity). In our example, the set of packets \textit{Packet x,2} is a block, and the switch aggregates them together. 
There are three main resources that the switch needs to manage: the cores (HPUs), the input buffer memory (i.e., the memory where the received packets are stored while being processed), and the working memory (i.e., the memory where the partially aggregated data is stored). The working memory is located in the L1 memory of the clusters, and the input buffers in the L2 packet memory (and copied in the L1 memory before starting the handler).  We show how these resources interact in Figure~\ref{fig:resources}, assuming a switch with 4 cores, processing the data shown in Figure~\ref{fig:blocks}. 

\subsection{Cores}\label{sec:design:cores}
\begin{figure}[ht]
    \centering
    \includegraphics[width=\columnwidth]{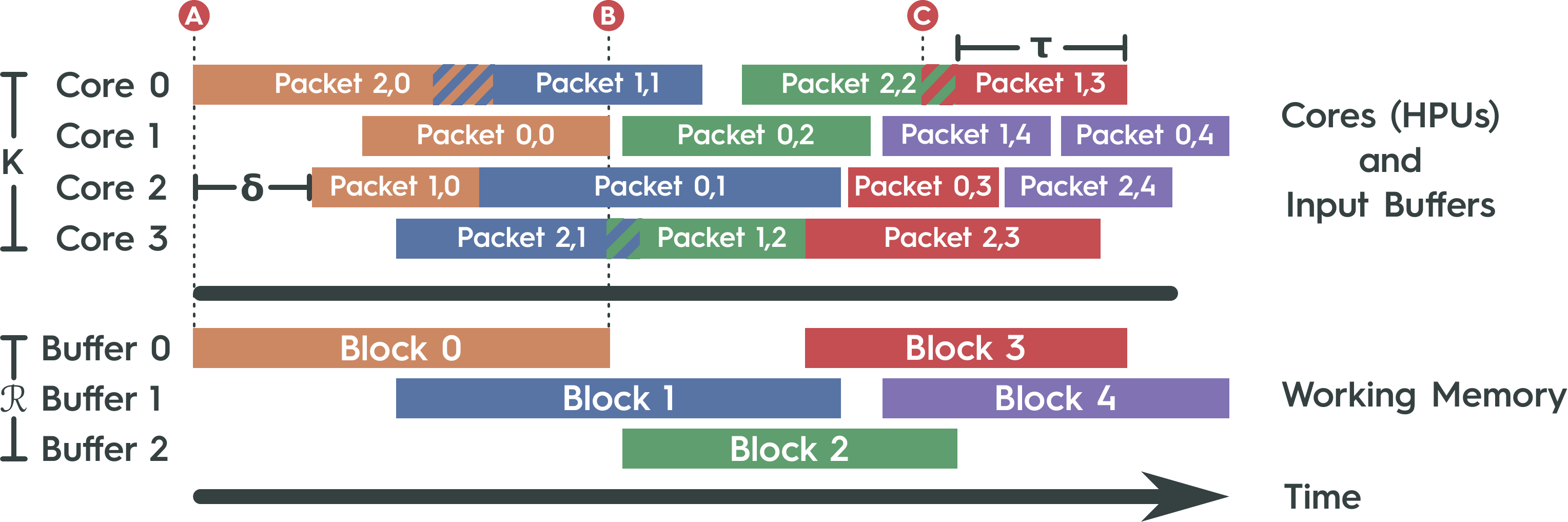}
    \caption{Utilization of input buffers, cores, and working memory during an in-network allreduce on a switch with 4 cores, processing the data shown in Figure~\ref{fig:blocks}.}
    \label{fig:resources}
\end{figure}

When a matching packet arrives at the processing unit of the switch, it is scheduled to one of the cores (we discuss scheduling more in detail in Section~\ref{sec:scheduling}). For example, in Figure~\ref{fig:resources}, \textit{Packet 2,0} is the first to arrive and it is scheduled to \textit{Core 0}. Because it is the first packet received for \textit{Block 0}, its content is simply copied in an available \textit{aggregation buffer} in the L1 memory (Figure~{\ref{fig:resources}}, {\includegraphics[scale=0.04,trim=0 30 0 0]{figs/marker_a.pdf})}. 
All the subsequent packets received for the same block are aggregated in the same buffer. For each block, the handler keeps a \textit{children} counter, incremented each time a packet of a block is received. The block can be considered fully aggregated when the counter is equal to the number of children in the reduction tree (specified when the handler was installed). In the example in Figure~{\ref{fig:resources}}, this happens at \includegraphics[scale=0.04,trim=0 30 0 0]{figs/marker_b.pdf} for \textit{Block 0}. The content of the aggregation buffer is then multicast to the children (or if the switch is not the root of the tree, sent to its parent), and the aggregation buffer is released, so that it could be used later by a subsequent block (in the example, by \textit{Block 3}). 

We assume that if a packet is lost, a timeout is triggered in the host, that retransmits the packet. To manage retransmissions, \flare can use a bitmap (with one bit per port) rather than a counter. When a packet of a block is received on a port, \flare checks the corresponding bit. If the bit is not set, the packet is aggregated and the bit is flipped. If the bit was already set, then the packet is a retransmitted packet, its content was already aggregated and it is not aggregated again. To simplify the exposition, in the following we assume refer to a \textit{children} counter rather than a bitmap.

The aggregation of the packets belonging to the same block can be done in different ways and has a direct impact on the maximum bandwidth that can be achieved by the switch when aggregating data. We denote with $\tau$ the service time of a core (i.e., the number of cycles it needs to aggregate a packet). Because we have $K$ cores (4 in the example), if the workload is evenly distributed across the cores (and we will show this is the case), the maximum bandwidth achievable by the switch is expressed as $\frac{K}{\tau}$. We assume the switch receives a packet every $\delta$ cycles and we can then express the bandwidth of the switch (in packets processed per cycle) as $\mathscr{B} = \min\big(\frac{K}{\tau}, \frac{1}{\delta}\big)$. $K$, $\tau$, and $\delta$ are determined by the specific switch and network design, and we show in Section~\ref{sec:parallelism} how to properly organize the computation so to minimize $\tau$.

\subsection{Input Buffers Memory}
While being processed, the packets occupy input buffers memory, for as long as the handler duration. Additionally, if there are no cores available for scheduling the packet, the packet waits in the input buffers memory until a core becomes available. In the example in Figure~{\ref{fig:resources}}, we denote this situation with a striped pattern in point \includegraphics[scale=0.04,trim=0 30 0 0]{figs/marker_c.pdf}, to indicate that the packet is sitting on a queue waiting to be scheduled. The input buffer occupancy is thus equal to the time the packet spends in the queue plus the time it spends while being processed. We denote the maximum size of these queues with $Q$. For example, \textit{Core 0} has $Q=0$ at  \includegraphics[scale=0.04,trim=0 30 0 0]{figs/marker_b.pdf}, and $Q=1$ at \includegraphics[scale=0.04,trim=0 30 0 0]{figs/marker_c.pdf}. We model and analyze this in detail in Section~\ref{sec:scheduling}.

\begin{figure*}[htpb]
   \begin{minipage}[b]{.59\textwidth}
    \centering
    \includegraphics[width=\columnwidth]{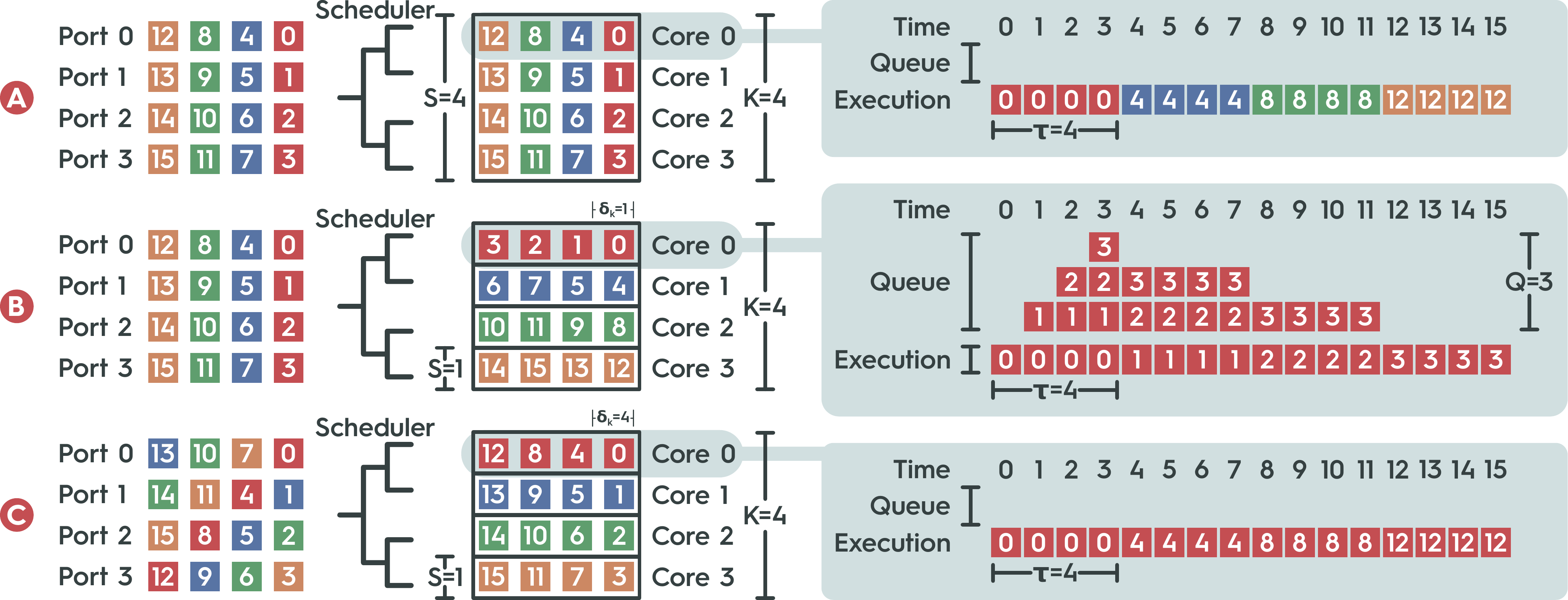}
    \captionof{figure}{Impact of intra-block interarrival time and hierarchical FCFS scheduling on packets memory occupancy.}
    \label{fig:scheduling}
  \end{minipage}
  \hfill
  \begin{minipage}[b]{.39\textwidth}
    \centering
    \footnotesize
     \begin{tabular}{|p{.5cm}|p{5cm}|} 
     \hline
     \textsc{Par.} & \textsc{Description} \\ 
     \hline\hline
     $K$ & Number of cores in the switch. \\ \hline
     $S$ & Number of cores in each scheduling subset. \\ \hline
     $P$ & Number of packets received for each block (this is equal to the number of children of the switch in the reduction tree). \\ \hline
     $\mathscr{Q}$ & Maximum number of packets in the switch. \\ \hline
     $\delta$ & Average interarrival time of packets. \\ \hline
     $\delta_c$ & Average interarrival time of packets belonging to the same block. \\ \hline
     $\delta_k$ & Average interarrival time of packets to a core (during a burst). \\ \hline
     $\tau$ & Average service time of a core. \\ \hline
     $M$ & Memory occupied by a block (number of elements). \\ \hline
      \end{tabular}
     \captionof{table}{Parameters used in the scheduling modeling.}
     \label{tab:schedule}
    \end{minipage}
\end{figure*}

\subsection{Working Memory}\label{sec:design:working}
We mentioned that the memory is partitioned among multiple allreduces, and in the example, we assume three buffers have been allocated to this specific reduction. For simplicity, we also assume each block is aggregated on a single buffer. However, multiple concurrent buffers per block could be used, for example, to reduce contention, and in Section~\ref{sec:parallelism} we describe different possibilities for organizing the working memory.
To avoid running out of memory, each host can have a number of \textit{``in-flight''} blocks not larger than the number of aggregation buffers assigned to that allreduce. In our example, the hosts send the fourth block only after the first block has been fully reduced and the buffer has been released. We can use Little's Law~\cite{littleslaw} to determine how many aggregation buffers should be allocated to each allreduce. Because we have $P$ packets per block (three in this case, equal to the number of hosts), the target bandwidth (in blocks per cycle), can be simply defined as $\mathscr{B}/P$. We define with $\mathscr{L}$ the latency (in cycles) to process a block (in our example, the time between \includegraphics[scale=0.04,trim=0 30 0 0]{figs/marker_a.pdf} and \includegraphics[scale=0.04,trim=0 30 0 0]{figs/marker_b.pdf} in Figure~{\ref{fig:resources}}), and with $M$ the number of buffers needed to aggregate a block. Then, each allreduce needs a working memory (in number of buffers) equal to $\mathscr{R} = M\frac{\mathscr{B}}{P}\mathscr{L}$.  

\section{Packets Scheduling and Input Buffers Occupancy}\label{sec:scheduling}
By default, packets are scheduled to the cores with a \textit{First Come First Serve} (FCFS) policy, so that they are evenly distributed across the cores. To simplify the exposition, we also assume that we size the system so that the \textit{interarrival time} to the processing unit (i.e., the time between the reception of two subsequent packets) is larger or equal than its \textit{service time} (i.e., the time between the sending of two subsequent packets). Under these conditions, on average packets will never be enqueued because they will always find an available core. In general, however, packets might be enqueued, waiting for a core to become available. When the queue is full, the packet is dropped or congestion is notified before filling the queue, depending on the specific network where the switch is integrated.

In PsPIN the L1 memory of the switch is partitioned across multiple clusters of cores (Figure~\ref{fig:switch}). This means that each of the aggregation buffers shown in Figure~\ref{fig:resources} is allocated on the L1 memory of a specific cluster. For example, if we assume to have two cores per cluster, and that \textit{Buffer 0} is allocated on the cluster of \textit{Cores 0} and \textit{1}, then the handler running on \textit{Core 2} would need to access a remote L1 memory. By doing so, it incurs a higher latency (up to 25x higher~\cite{pspin}) compared to accessing its local L1 memory. 

To only have local L1 accesses and improve performance, we restrict the processing of packets belonging to the same block to a subset of cores (located on the same cluster). To keep the workload balanced among all the available cores, thus still guaranteeing line-rate processing, we adopt hierarchical FCFS scheduling. We assign packets belonging to the same block with an FCFS policy to the same subset of cores, and different blocks to different subsets\footnote{The identifier of the block can be carried in the packet as an IP optional header, processed by the \textit{parser} and communicated to the packet scheduler.}. Even if in the long run this evenly distributes the packets to the cores, it might generate short bursts of packets directed to the same core(s), that need to be enqueued, thus increasing the occupancy of the packets buffers memory. We analyze this more in detail in Section~\ref{sec:scheduling}.

To understand the impact of scheduling decisions on the input buffer occupancy,  we illustrate in Figure~\ref{fig:scheduling} three different scenarios (\includegraphics[scale=0.04,trim=0 30 0 0]{figs/marker_a.pdf}, \includegraphics[scale=0.04,trim=0 30 0 0]{figs/marker_b.pdf}, and \includegraphics[scale=0.04,trim=0 30 0 0]{figs/marker_c.pdf}), and we report in Table~\ref{tab:schedule} some variables we use in our analysis. In scenario \includegraphics[scale=0.04,trim=0 30 0 0]{figs/marker_a.pdf} packets are received from four ports (we suppose there is a host attached to each of them), and packets with the same color belong to the same block and must be aggregated together. The number inside each packet represents the time when it is received by the switch. Because each packet might spend some time in the queue this is, in general, different from the time when a core starts to process it. Due to congestion and unbalance, arrival times are not so nicely aligned, and there might be gaps between the reception of subsequent packets. However, for illustrative purposes, we assume the simplest case where all packets are received at a steady and constant rate of one packet per second. 
We assume that the switch has $K$ cores (4 in this case), $P$ ports, that it receives one packet every $\delta$ seconds (1 in this case), and that each core has a service time $\tau$ (4 seconds in the example). Because the service time of the switch is $\frac{\tau}{K} = \delta$, if packets are evenly distributed the switch can process the packets at line rate.
We also define $\delta_{c}$ as the interarrival time of the packets \textit{within a block} ($\delta_{c} = 1$ on scenario \includegraphics[scale=0.04,trim=0 30 0 0]{figs/marker_a.pdf}). On the right part of the figure we show a detail of what happens in \textit{Core 0}. Because it receives a packet every 4 seconds, and its service time is 4 seconds, packets are never enqueued.

On scenario \includegraphics[scale=0.04,trim=0 30 0 0]{figs/marker_b.pdf} of Figure~{\ref{fig:scheduling}}, packets belonging to the same block, are instead assigned to a subset of cores (each containing one core, i.e., $S=1$). Each core now does not receive a steady flow of packets, but bursts of packets at regular intervals. Whereas on scenario \includegraphics[scale=0.04,trim=0 30 0 0]{figs/marker_a.pdf} each core receives one packet every 4 seconds, on scenario \includegraphics[scale=0.04,trim=0 30 0 0]{figs/marker_b.pdf} each core receives 4 packets in 4 seconds, and then nothing for 12 seconds. Even if in both \includegraphics[scale=0.04,trim=0 30 0 0]{figs/marker_a.pdf} and \includegraphics[scale=0.04,trim=0 30 0 0]{figs/marker_b.pdf} each core receives on average 4 packets every 16 seconds, on scenario \includegraphics[scale=0.04,trim=0 30 0 0]{figs/marker_b.pdf} the bursts build up queues in front of each core. On the right part of the figure, we show in detail what happens in \textit{Core 0}. The core receives 4 packets in 4 second, thus building a queue, that will eventually be completely absorbed before the beginning of the next burst (because we assumed to size the system to process packets, on average, at line rate). However, these queues increased the time a packet spent in the switch and, thus, the input buffer occupancy.

Moreover, the intensity of the bursts (and thus the queue length) does not only depend on the size of the subsets $S$, but also on $\delta_c$. For example, in scenario \includegraphics[scale=0.04,trim=0 30 0 0]{figs/marker_c.pdf} of Figure~{\ref{fig:scheduling}} we show the same scenario as \includegraphics[scale=0.04,trim=0 30 0 0]{figs/marker_b.pdf}, but now with $\delta_c=4$, that implies that packets belonging to the same block will arrive four seconds apart one from on another. By comparing \includegraphics[scale=0.04,trim=0 30 0 0]{figs/marker_c.pdf} with \includegraphics[scale=0.04,trim=0 30 0 0]{figs/marker_b.pdf} we can observe that although packets arrive at the same rate (one packet per second in both cases), and the scheduler assigns the packets to the same subset of cores, on scenario \includegraphics[scale=0.04,trim=0 30 0 0]{figs/marker_c.pdf} packets of the same block arrive at each core at a slower rate and are never enqueued, as we can observe in the detail of \textit{Core 0}. In this scenario, we obtain the same locality as in scenario \includegraphics[scale=0.04,trim=0 30 0 0]{figs/marker_b.pdf} (packets belonging to the same block are assigned to the same subset of cores), and the minimal input buffer occupancy as in scenario \includegraphics[scale=0.04,trim=0 30 0 0]{figs/marker_a.pdf}.

$\delta_c$ depends on non-controllable factors such as application imbalance~\cite{li2020taming,imbalance}, network noise~\cite{10.1145/3295500.3356196, allreduce, htornnoise,Chunduri:2017:RVX:3126908.3126926,1526010}, OS noise~\cite{osnoise,6012894}, and network contention~\cite{ibm:tm,fattree:sc18,doi:10.1142/S0129626409000419,bully,8425264,sc18}, but also on some controllable factors, such as the order the packets are sent by the hosts. In this work, we propose a solution called \textit{staggered sending}, that consists in having each host sending the packets in a different order so that, on average, packets belonging to the same block can be scheduled to a specific subset of cores, while not increasing the size of the queues and, thus, the input buffer occupancy. Moreover, as we show in Section~\ref{sec:parallelization:single-buf}, \textit{staggered sending} is also helpful in reducing contention on the shared aggregation buffer. In general, the maximum $\delta_c$ we can induce with staggered sending depends on the number of blocks to be sent. In the example in scenario \includegraphics[scale=0.04,trim=0 30 0 0]{figs/marker_c.pdf}, if we would have only 2 blocks, the $\delta_c$ would be half of that we have when having 4 blocks. In general, we have $\delta \leq \delta_c \leq  \delta \frac{Z}{N}$.  

\paragraph{Input buffer occupancy}
Because we are considering bursty arrival times at the cores, we can't use Little's Law~\cite{littleslaw} to compute the average number of packets in the switch, because it would consider an average case, and would not capture differences between the three scenarios. Instead, we first compute the interarrival time of the packets in a burst to a specific core, that we indicate with $\delta_k$. Packets arrive to a subset of $S$ cores with an interarrival time $\delta_c$, and thus to each of the cores in the subset with interarrival $\delta_k = S\delta_c$. Because we on the long run packers are evenly distributed across the cores, this can never be higher than $K\delta$, and thus we have $\delta_k = \min(S\delta_c, K\delta)$. Because $P$ packets are received for each block, a burst can contain up to $\frac{P}{S}$ packets, and it takes $\delta_k \frac{P}{S}$ cycles to completely receive the burst. By that time, $\frac{\delta_kP}{\tau S}$ packets have been processed by the core and removed from the queue. Accordingly, we can express the maximum queue length as $Q=\frac{\delta_kP}{\delta_kS} - \frac{\delta_kP}{\tau S} = \frac{P}{S}\big(1-\frac{\delta_k}{\tau}\big)$, and the maximum number of packets in the switch (including those being currently processed by each core) as:
\begin{equation}\label{eq:packets_in_switch}
\mathscr{Q} = (Q+1) K = \frac{PK}{S}\bigg(1-\frac{\delta_k}{\tau}\bigg) + K
\end{equation}

This equation shows the relationship between the scheduling decision and the input buffer occupancy (e.g., the smaller $S$, the higher the input buffer occupancy). It can also be used to compute the latency $\mathscr{L}$ to process a block and, thus, the working memory occupancy (Section~\ref{sec:design:working}). Indeed, the latency can be computed as the time the switch waits for all the packets of the block to be received ($(P-1)\delta_c$), plus the time needed for processing the last packet. This includes both the time spent aggregating the packet and the time spent in the queue. In the worst case, a packet spends $Q\tau$ cycles in the queue, and the latency is equal to $\mathscr{L} = (P-1)\delta_c + (Q+1)\tau$. We use the queueing time and the latency models for estimating the input buffer and working memory occupancy in Section~\ref{sec:parallelism}.





\section{Parallelism and memory organization}\label{sec:parallelism}
             


We now describe the design of different aggregation approaches that can be used to reduce a block of $P$ packets in \flare. We consider different designs: aggregation on a single memory buffer shared by all the packets of a block (Section~\ref{sec:parallelization:single-buf}), aggregation on multiple buffers (Section~\ref{sec:parallelization:multi-buf}), and asynchronous tree aggregation (Section~\ref{sec:parallelization:tree}). We then analyze the different tradeoffs between these alternatives in Section~\ref{sec:evaluation:types}.

When analyzing the different aggregation approaches we model the service time of a core $\tau$ (Section~\ref{sec:design:cores}). This can then be used to model the size of the working memory (Section~\ref{sec:design:working}); and the input buffer occupancy (Equation~\ref{eq:packets_in_switch}). To model the working memory size, we also need to model $M$, i.e., the number of buffers used for each block. In our modeling, we assume to have 1KiB packets containing 256 floating-point values. We measured the time required to aggregate a packet in the aggregation buffer by using the PsPIN cycle-accurate simulator. On average, a core of the PsPIN unit needs four cycles to sum two 4-bytes floating point values and to store the result back in the aggregation buffer. Because the unit is clocked at 1GHz (Section~{\ref{sec:switch}}), the time required to process a packet is 1ns per byte circa, and we use this information to model $\tau$.

\subsection[]{Aggregation using a single buffer}\label{sec:parallelization:single-buf}
\begin{figure}[htpb]
    \centering
    \includegraphics[width=\columnwidth]{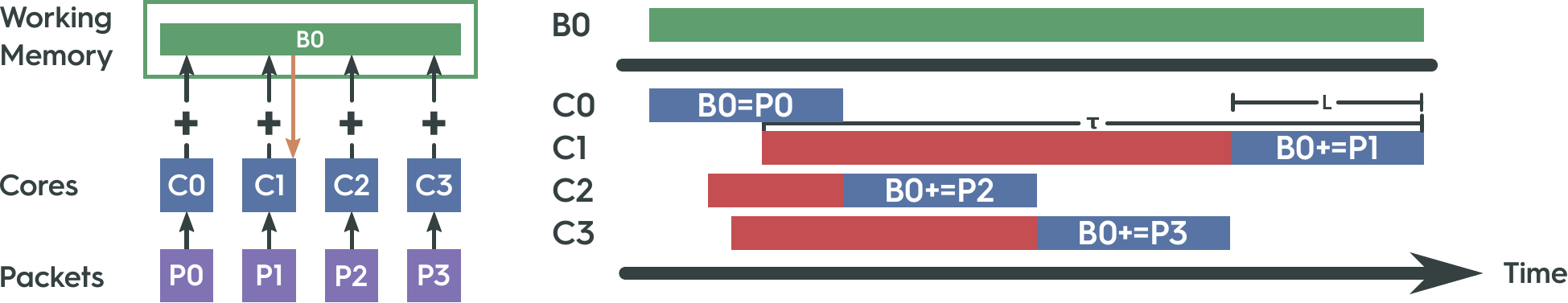}
    \caption{Single buffer aggregation.}
    \label{fig:single_buffer}
\end{figure}

The first approach we propose is the most straightforward one, where all the packets of the same block are accumulated in the same working memory buffer, as shown in Figure~\ref{fig:single_buffer}. We show on the left the packets processed by each core, and how the cores access the buffer. In this case, cores \textit{C0}, \textit{C2}, and \textit{C3} just sum the content of their packets into the buffer. Cores \textit{C1} also reads back the fully aggregated result, that will then send on a packet over the network. On the right, we show the timeline of these operations. For the cores, the boxes represent the duration of the handlers, whereas for the buffer we depict for how long the buffer has been used.

We assume the most general case where the concurrent aggregation of multiple packets in a shared aggregation buffer is executed in a critical section. Indeed, even if in principle we could use atomic operations~\cite{atun}, the computation would still be affected by severe contention and/or performance overhead~\cite{atomics,fireforget}. Moreover, by assuming a critical section we cover the cases where the user function cannot be executed by using atomic operations, and the aggregation of sparse data that, as we will show in Section~\ref{sec:sparsity}, requires more complex processing and in most cases needs to be executed anyhow in a mutually exclusive way.

To avoid expensive context switches, PsPIN handlers are never suspended and terminate only after the packet has been processed. Thus, if waiting to enter a critical section, they will actively consume compute cycles on one of the cores. For example, when a packet arrives at \textit{C2}, the core waits for the buffer to be released by \textit{C0} (we denote this waiting time with a red box), and only then it starts the aggregation. We define with $L$ the number of cycles needed to aggregate a block after the handler enters the critical section (i.e., in general, $L \leq \tau$). The probability that two handlers need to access the same aggregation buffer concurrently depends on $\delta_c$, i.e., the interarrival time of the packets belonging to the same block. If $\delta_c \geq L$, on average there is never more than one packet of the same block processed concurrently. This is also true if all the packets of a block are executed by the same core (i.e., $S = 1$). However, for $S=1$ the input buffers occupancy would significantly increase, as discussed in Section~\ref{sec:scheduling}.

To estimate the cost of contention, we assume to have $C$ cores in each cluster. Because we schedule the packets belonging to the same block to a subset of $C$ cores, in the worst case we have $C$ concurrent packets of the same block that need to access the same aggregation buffer in the working memory. If a handler needs $L$ cycles to aggregate the packet (excluding the time spent waiting to enter the critical section), in the worst case, the first handler to be executed needs $L$ cycles to aggregate the packet, the second handler $2L$ cycles, the third handler $3L$ cycles, etc\ldots In general, the average service time of a core can thus be expressed as $\tau = \frac{\sum_{i=1}^C iL}{C} = \frac{L(C-1)}{2}$.


We then have:

\begin{equation}\label{eq:single_buffer:work}
  \tau \leq \begin{cases}
    L , & \text{$S=1$ or $\delta_c \geq L$}.\\
    \frac{L(C-1)}{2}, & \text{otherwise}.\\
  \end{cases}
\end{equation}

We minimize $\tau$ for $S=1$ or $\delta_c \geq L$. As discussed in Section~\ref{sec:scheduling}, we can change $S$ by restricting the execution of the packets of the same block to a subset of $S$ cores. According to equation~\ref{eq:packets_in_switch}, by reducing $\tau$ we would decrease the input buffers occupancy. However, this increase might be canceled out because we are also decreasing $S$. On the other hand, by increasing $\delta_c$ we always decrease the input buffer occupancy, but we increase the latency required for processing a block and, thus, the working memory occupancy. 

Before showing the effect of these decisions on both bandwidth and memory occupancy, we observe that we can increase $\delta_c$ by using \textit{staggered sending}, as discussed in Section~\ref{sec:scheduling}. In a nutshell, this means that to avoid contention there should never be two cores working on two packets of the same block. This can only happen if the data to be reduced is large enough to have, at any time, one block per core. For our switch, this can only be guaranteed if the hosts use staggered sending and if the size of the data to be reduced is larger than 512KiB. Because small-size allreduces are a significant fraction of the allreduce traffic~\cite{mpi-characterization}, we propose in the next sections some alternative approaches to address such cases. Moreover, because we only need one buffer per block, we have $M=1$.

\subsubsection{Main insights}
\begin{figure}[htpb]
     \centering
     \includegraphics[width=\columnwidth]{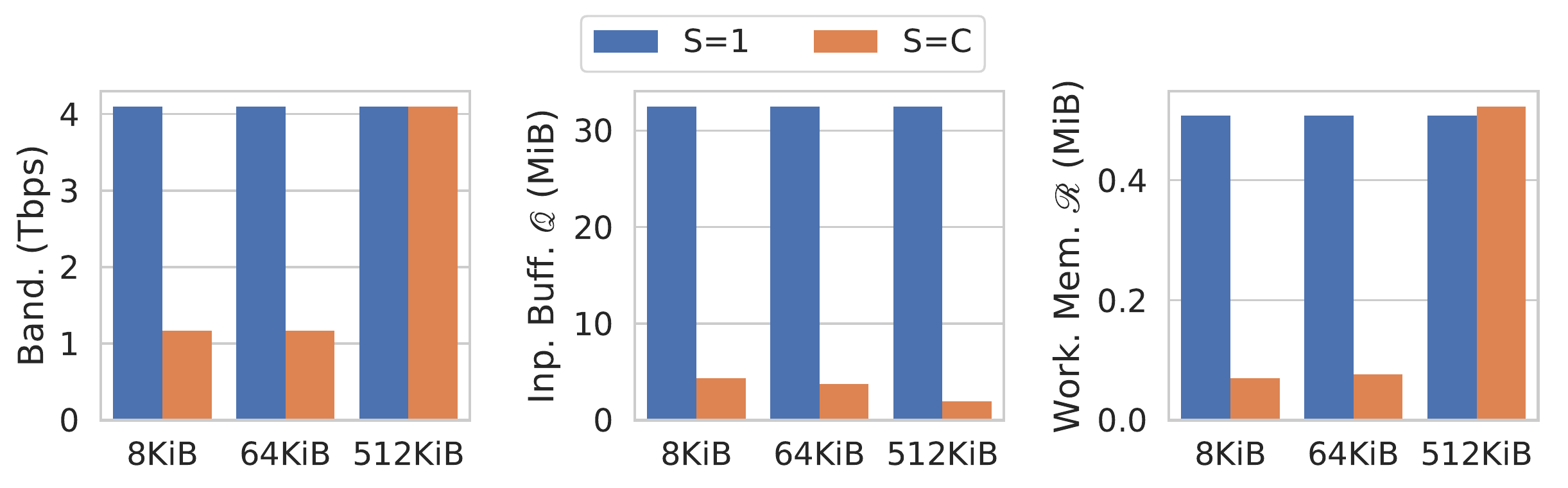}
    \caption{Bandwidth and memory occupancy of single buffer aggregation.}
    \label{fig:results:single-buffer}
\end{figure}

Figure~\ref{fig:results:single-buffer} illustrates the modeled bandwidth, memory occupancy of the input buffers ($\mathscr{Q}$), and occupancy of the working memory ($\mathscr{R}$). On the x-axis, we have different values of $S$. We consider restricting the execution of packets belonging to the same block to 1 core, and $C$ cores (i.e., all the cores of a cluster). 

First, we observe that for small messages (where we can't sufficiently increase $\delta_c$) there is a large performance drop when $S=C$. However, having $S=1$ significantly increases the memory occupancy. For larger messages, we observe good performance also when scheduling the packets on $C$ cores of the same cluster, thus decreasing the queue length and the occupancy of the input buffers memory. The occupancy of the working memory is negligible and around 512KiB. For large enough packets and $S=C$, the maximum total memory occupancy is 2MiB for a single reduction. In general, scheduling the packets belonging to the same reduction block to a single core significantly increases the memory occupancy and, for this reason, in the following we design alternative solutions that can achieve a higher bandwidth and a lower memory occupancy for smaller allreduces.

\subsection[]{Aggregation on multiple buffers}\label{sec:parallelization:multi-buf}

\begin{figure}[h!]
    \centering
    \includegraphics[width=\columnwidth]{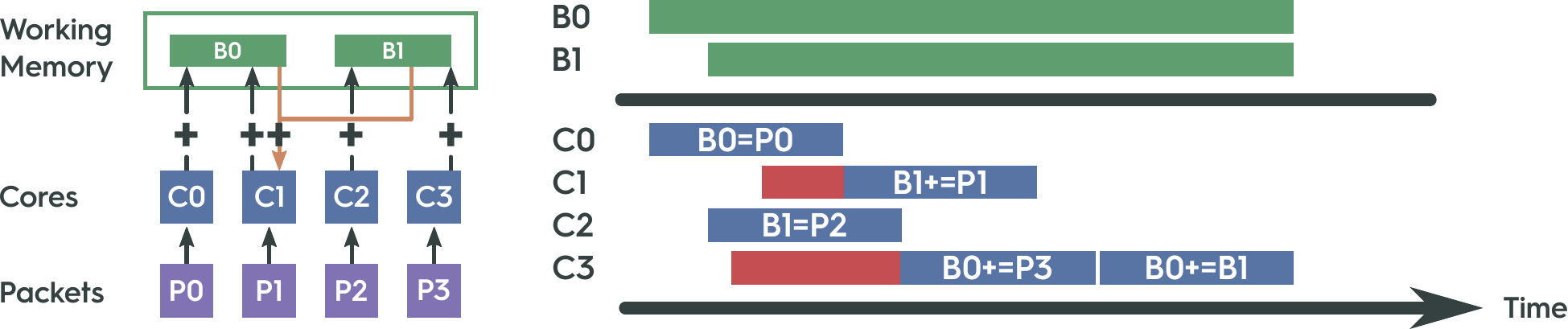}
    \caption{Multiple buffer aggregation.}
    \label{fig:multi_buffer}
\end{figure}

An extension to the previous aggregation design consists in having $B$ aggregation buffers for each block. In Figure~\ref{fig:multi_buffer} we show an example with two buffers. When a handler is executed, it takes whichever of those buffers is not currently used by any other handler. In the example, differently from the single buffer case, when \textit{C2} receives the packet, it finds \textit{B0} already being used by \textit{C0}, and thus aggregates the packet on \textit{B1}. If no buffer is available, the handler needs to wait to enter a critical section, as for the single buffer aggregation (for example, this is what happens to both \textit{C1} and \textit{C3}). Because the aggregated data is now distributed over $B$ buffers, the last handler to be executed for the block needs to aggregate together the partially aggregated data contained in the remaining $B-1$ buffers. In the example, the handler running on \textit{C1} is the last one to be executed and, aggregates the content of its packet with the content of \textit{B0}, and then of \textit{B1}. 

$\tau$ can be computed starting from Equation~\ref{eq:single_buffer:work}, by replacing $\delta_c$ with $B\delta_c$. Indeed, when using $B$ buffers, the probability that two running handlers need to access the same buffer decreases proportionally with $B$. Compared to the single buffer aggregation, this solution is less affected by contention for shorter interarrival times (and, thus, for smaller data). Moreover, the last handler needs to aggregate the data from the other $B-1$ buffers, and this costs $(B-1)L$ additional cycles. Because $B$ buffers are stored for each block, we have $M=B$.

\subsubsection{Main insights}
This aggregation algorithm increases the concurrency and the bandwidth at the cost of higher memory occupancy. However, using too many buffers reduces the bandwidth, due to the extra cost required for sequentially aggregating the partially reduced data of each buffer. We analyze the tradeoffs between bandwidth and memory occupancy in Section~\ref{sec:evaluation:types}.

\subsection[]{Tree aggregation}\label{sec:parallelization:tree}

\begin{figure}[ht]
    \centering
    \includegraphics[width=\columnwidth]{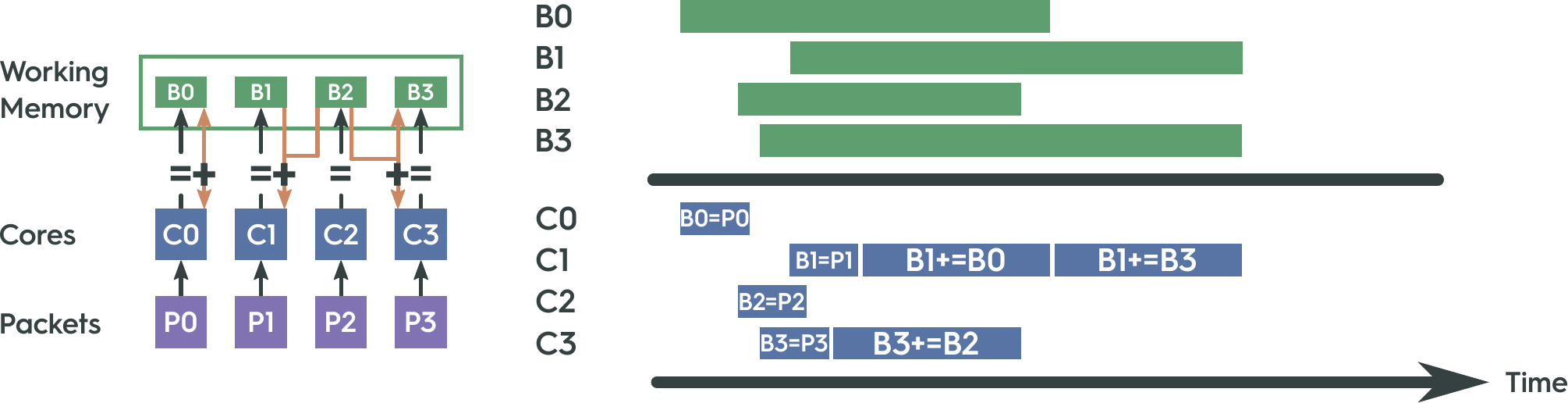}
    \caption{Tree aggregation.}
    \label{fig:tree}
\end{figure}

The main drawback of the multiple buffer aggregation, is the sequential aggregation of the data contained in each of the aggregation buffers. To avoid that cost, we can organize the computation in a tree-like way. We illustrate this idea in Figure~\ref{fig:tree}. When a core receives a packet, it allocates a buffer and copies the data into it. After copying the data, a core might also aggregate the data contained in some of the other buffers. The idea is to organize the aggregation of those buffers in a pre-defined tree-like way. For example, the data contained in \textit{B0} can only be aggregated with \textit{B1}, and the result stored back in \textit{B1} (after this aggregation step, \textit{B0} can be deallocated). Similarly, the data contained in \textit{B2} can only be aggregated with \textit{B3} and stored back in \textit{B3} (and \textit{B2} deallocated). When \textit{B1} contains the result of \textit{B0+B1}, and \textit{B3} the result of \textit{B2+B3}, the final aggregation of \textit{B1+B3} can be computed.

To completely avoid contention on shared aggregation buffers (independently from $\delta_c$ and, thus, from the message size) the computation on the next level of the tree is carried only if a core finds available data in both buffers. For example, when \textit{C0} receives the packet, it copies it in \textit{B0}, and could aggregate \textit{B0} with \textit{B1}. However, there is no available data yet in \textit{B1}, and the handler is terminated. Instead, when \textit{C1} processes its packet, it finds the data in \textit{B0} and carries on the computation of the next level of the tree. It also finds the result of \textit{B2+B3} in \textit{B3}, and thus also computes \textit{B1+B3}, and sends the completely aggregated result on the network. To guarantee the reproducibility of floating-point summation, a packet coming from a port $i$ is always stored in a buffer $j$. This forces the tree to always have the same structure and guarantees the reproducibility of floating-point summation across different runs. 

To compute the service time, we assume that the cost of copying the data is negligible. For example, in PsPIN this can be done through the DMA engine, at the cost of 64 cycles instead of the 1024 cycles needed for the aggregation. A total of $P-1$ aggregations are executed, and we can compute the average number of cycles per packet as $\tau = \frac{(P-1)L}{P}$. To compute the memory consumption, we observe that each time two buffers are aggregated, one of them is discarded. Because $P-1$ aggregations are performed and they are arranged in $\log P$ levels in the tree, on average we have $M = \frac{P-1}{\log P}$ active buffers for each block. 

\subsubsection{Main insights} Differently from single and multiple buffers aggregation, in this case, the handlers never waste cycles waiting for entering a critical section, and this design achieves the optimal bandwidth independently from $\delta_c$ and, thus, from the data size. However, it requires more buffers per block, increasing the average memory occupancy. We analyze the bandwidth, latency, and memory occupancy of this design in Section~\ref{sec:evaluation:types}.

\subsection{Evaluation}\label{sec:evaluation:types}
\begin{figure}[ht]
    \centering
    \includegraphics[width=\columnwidth]{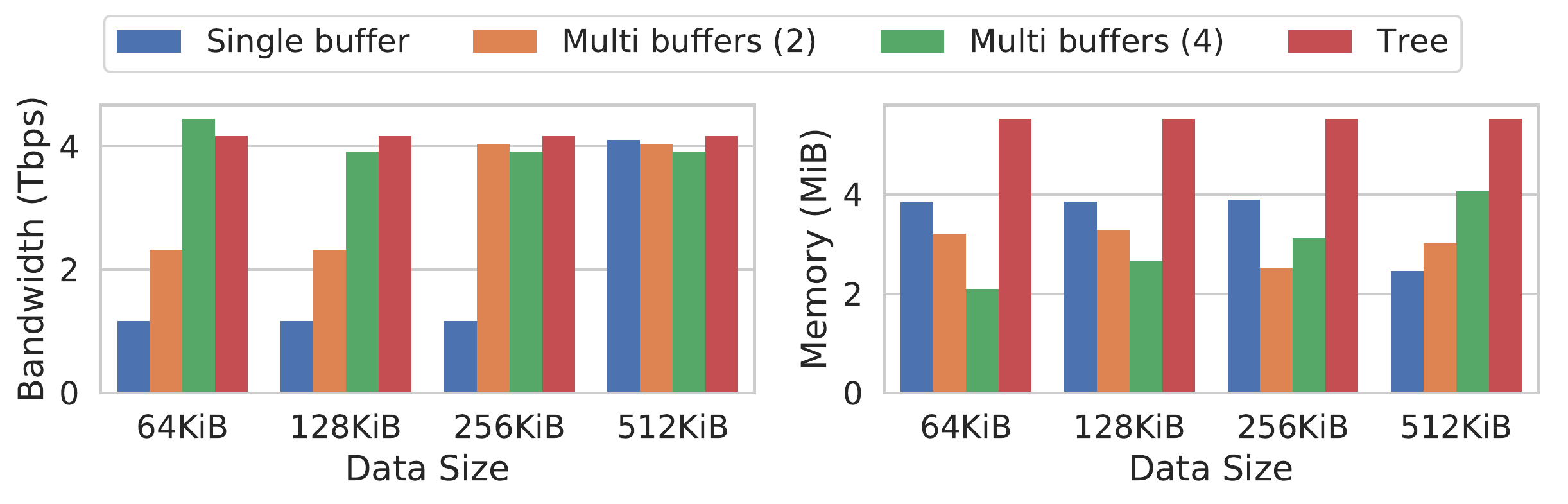}
    \caption{Modeled bandwidth and memory occupancy for $S=C$ and different data sizes.}
    \label{fig:res}
\end{figure}

In Figure~\ref{fig:res} we show the modeled maximum bandwidth and the memory occupancy, for $S=C$ and different data sizes. We observe that the only best performing algorithm on data smaller than 128KiB is the tree aggregation. When increasing the data size, multi-buffers aggregation catches up, and the higher the number of buffers, the higher the bandwidth for smaller messages. Eventually, for data larger than 512KiB, single buffer aggregation catches up with the other solutions. 

Regarding the memory occupancy, it is worth remarking that, in some cases, using more buffers decreases the memory occupancy, because the performance is higher and those buffers are used for a shorter time. To optimize both compute and memory resources, \flare uses single buffer aggregation if the size of the data to be reduced is larger than 512KiB, multi buffers with 4 buffers if larger than 256KiB, with 2 buffers if larger than 128KiB, and tree aggregation otherwise. When reproducibility of floating-point summation is required, \flare always uses tree aggregation.

\begin{figure}[ht]
    \centering
    \includegraphics[width=\columnwidth]{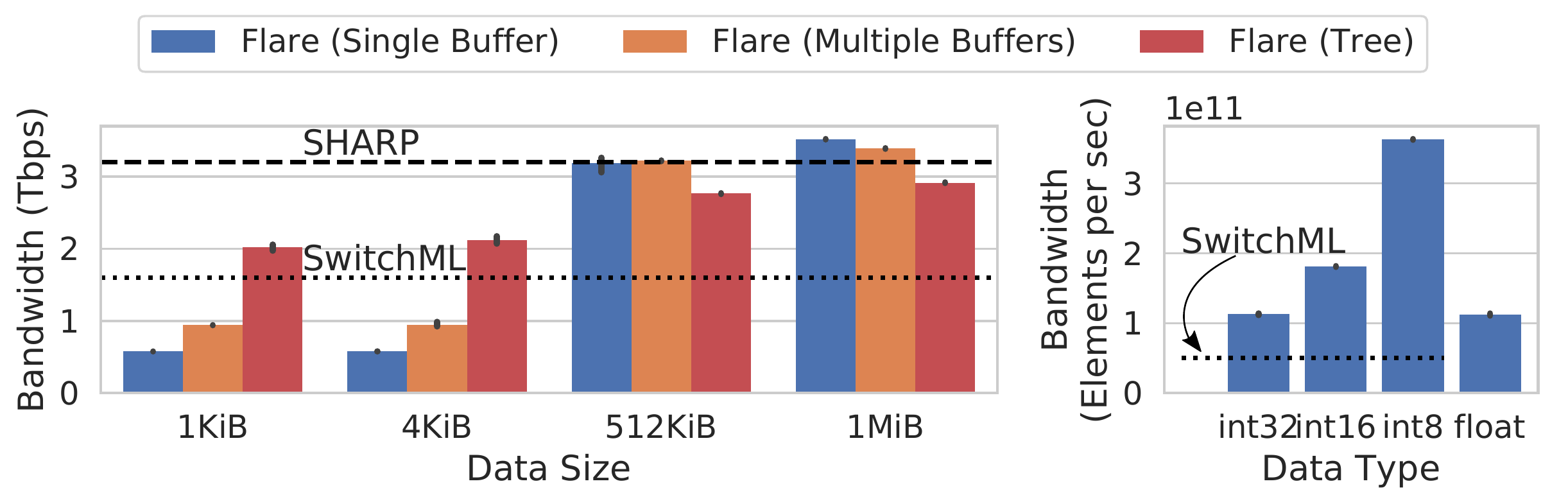}
    \caption{Bandwidth for different data size and data types.}
    \label{fig:simulated_data_types}
\end{figure}

We implemented the different aggregation algorithms in the PsPIN cycle-accurate simulator~\cite{pspin}. To simulate delays in the hosts sending the data and in the network, we generate packets with a random and exponentially distributed arrival rate. The actual PsPIN implementation~\cite{pspin} only simulates 4 clusters. Because the clusters are organized in a shared-nothing configuration, we scale the results linearly with the number of deployed clusters. First, we report in Figure~\ref{fig:simulated_data_types} the maximum bandwidth that the switch can achieve for the aggregation of 32-bits integers vectors of different size $Z$. We compare \flare to two baselines: SwitchML~\cite{switchml} and SHARP~\cite{sharp,sharp2}. SwitchML runs on programmable Tofino switches~\cite{tofino}, can only process integer elements, and achieves a maximum bandwidth of 1.6Tbps~\cite{switchml}. SHARP is a solution running on Mellanox's fixed-function switches~\cite{quantum}, and can process floating-point elements. Switches supporting SHARP have 40 ports at 200Gbps. However, to the best of our knowledge, the best available known data for SHARP (for a single switch) shows a 3.2Tbps bandwidth~\cite{sharp2} (32 ports at 100Gbps), and we use this as a reference.

We observe that, for small data, only tree aggregation provides higher bandwidth than SwitchML. Indeed, single and multi-buffers aggregation cannot exploit staggered sending for small data, and experience contention when accessing the aggregation buffers. Moreover, for small data, we are showing a \textit{``cold start''} case, where the handlers were not loaded yet in the instruction cache. For larger data, single buffer aggregation efficiently exploits staggered sending, achieving a higher bandwidth compared to multi-buffer and tree aggregation. Indeed, both multi-buffer and tree aggregation have some additional overhead caused by the management of multiple buffers. We also observe that even \flare achieves a higher bandwidth compared to SwitchML and SHARP, while also capable of running arbitrary computations on network packets. As we show in Section~\ref{sec:evaluation:sparse} this leads, for example, to significant performance improvements when dealing with sparse data.

We also report in Figure~\ref{fig:simulated_data_types} the bandwidth (in elements aggregated per second), for different data types, for the aggregation of $1MiB$ data. We consider 32-bits, 16-bits, and 8-bits integers, and floating-point elements (not supported by SwitchML and by existing programmable switches in general). Programmable switches can only process a fixed number of elements per packet. Processing more elements per packet would require more \textit{recirculations}, thus decreasing the bandwidth accordingly. On the other hand, in \flare the HPUs of the PsPIN unit use vectorization and can aggregate, for example, two \texttt{int16} elements in a single cycle. This leads to an increase in the number of elements aggregated per second when sending elements with a smaller data type. However, differently from SHARP, \flare currently does not support the aggregation of double-precision floating-point elements.

\section{Sparse data}\label{sec:sparsity}
\begin{figure}[ht]
    \centering
    \includegraphics[width=\columnwidth]{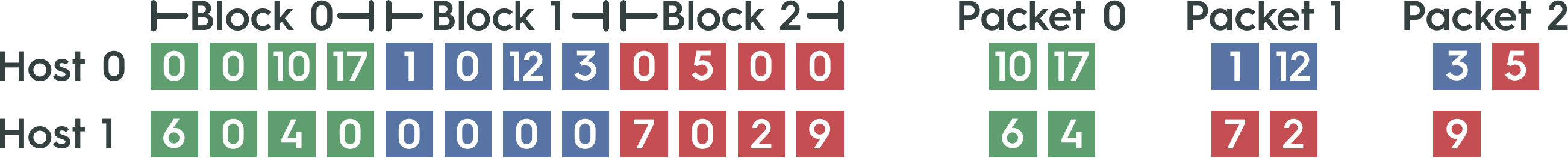}
    \caption{Packetization of sparse data. Packets also carry the position of the elements (not shown).}
    \label{fig:sparse_splitting}
\end{figure}

To reduce the network traffic and increase the performance, applications dealing with sparse data might send only the non-zero elements and their positions. In this case, each host splits the data into blocks so that, on average, each block contains a number of non-zero elements that would fit in a network packet. This implies that different hosts might have a different number of elements in each block, as shown in Figure~\ref{fig:sparse_splitting}. We show both the partitioning of the data in reduction blocks and the elements carried by each packet (packets also carry the position of each element inside the block, not shown in the figure for the sake of clarity). The number inside each box represents the value of that specific element. For simplicity, we assume that (due to the MTU) the hosts can send at most two elements per packet and that the data \textit{density} (defined as the average percentage of non-zero elements in each reduction block) is $50\%$. Thus, on average, there will be 2 non-zero elements every 4 elements, and thus we set the span of the block to 4 elements. 

However, whereas this is true on average, this introduces the following additional challenges compared to the dense case:
\paragraph{\textbf{Multiple blocks per packet}} Elements that belongs to two different blocks would be sent in the same packet (like the elements \includegraphics[scale=.3,trim=0 3 0 0]{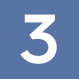} and \includegraphics[scale=.3,trim=0 3 0 0]{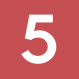} in Figure~{\ref{fig:sparse_splitting}}).  This implies that the handler must check, for each element in the packet, the block it belongs to (to aggregate it in the correct aggregation buffer), for example by dividing its index by the number of blocks. This introduces some additional computation for each element. To avoid this additional overhead, we send the block identifier in the packet header (as for the dense case), and we force the hosts to never send multiple blocks in the same packet, and rather to send a packet even if it is not full (in our example, a packet with only \includegraphics[scale=.3,trim=0 3 0 0]{figs/sparse_3.pdf}). 
\paragraph{\textbf{Block split}} Elements belonging to the same block could be split in two different packets (e.g., elements \includegraphics[scale=.3,trim=0 3 0 0]{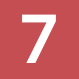} \includegraphics[scale=.3,trim=0 3 0 0]{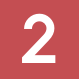} and \includegraphics[scale=.3,trim=0 3 0 0]{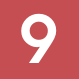} in Figure~{\ref{fig:sparse_splitting}}). For dense reductions, it was sufficient to keep a \textit{children} counter, incremented when a packet was received from a child, and consider the reduction of the block completed when the counter was equal to the number of children. In the sparse case, a node could receive multiple packets for the same block, and this approach is not sufficient anymore. To address this issue each host sends, in the last packet of a block, a counter representing the number of packets composing that block. For example, because node 1 splits block 2 in two packets, it sends the packet carrying element \includegraphics[scale=.3,trim=0 3 0 0]{figs/sparse_9.pdf} with the counter set to $2$. The handlers keep for each block an additional \textit{shard} counter for each of the $P$ hosts. The \textit{shard} counter is incremented every time a packet for that block is received from that port. When the counter is equal to the value carried in the last packet of the block, then all the packets for that block and from that specific port have been received, and the \textit{children} counter can be increased.

\paragraph{\textbf{Empty blocks}}
In some cases, we could have all-zero blocks (for example, block 1 sent by node 1). In those cases, we still send a packet with no elements (and with only the header with the identifier of the block), so that the switch can increase the children counter nevertheless. If we assume uniformly distributed zero values, this however should rarely happen. 


Another difference compared to the dense case is the design of the data structure that holds the partially aggregated data. \flare still aggregates separate blocks in different buffers. However, whereas for the dense case the aggregation buffer has size $N$, in the sparse case aggregating two buffers of $N$ elements leads, in general, to more than $N$ aggregated elements (because the non-zero elements in the two buffers might not perfectly overlap). For example, in Figure~{\ref{fig:sparse_splitting}}, aggregating the data in the packet \includegraphics[scale=.3,trim=0 3 0 0]{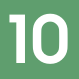} \includegraphics[scale=.3,trim=0 3 0 0]{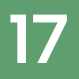} with the data in the packet \includegraphics[scale=.3,trim=0 3 0 0]{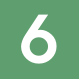} \includegraphics[scale=.3,trim=0 3 0 0]{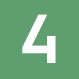} leads to 3 elements (\includegraphics[scale=.3,trim=0 3 0 0]{figs/sparse_6.pdf} \includegraphics[scale=.3,trim=0 3 0 0]{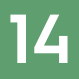} \includegraphics[scale=.3,trim=0 3 0 0]{figs/sparse_17.pdf}).

For sparse data, \flare stores the data and the indexes in a hash table. To avoid expensive collision resolution, when there is a collision, the colliding element is put in a \textit{spill buffer}. When the spill buffer is full, the spilled data is immediately sent to the next switch (or to the hosts). For highly sparse data, this solution reduces memory consumption compared to storing it in a dense format. For denser data, this however increases both the memory occupancy (because indexes must be explicitly stored), the latency when aggregating the data, and the network traffic (because more data spills).

For denser data, \flare uses a contiguous memory buffer of the size of the block. From a computational perspective, this is the design with the lowest latency, because the handler simply needs to store the element in a specific position. However, when the reduction is completed, the buffer needs to be entirely scanned and only the non-zero elements inserted in the packet. Moreover, the memory consumption will be equal to that of the dense case, thus much higher than the optimal if the buffer contains many zeros.  

In general, sparse data get denser after each aggregation and, when aggregating data on an in-network reduction tree, the data get denser while traveling from the hosts to the root of the tree. For this reason, \flare stores the data in hash tables in the leaves switches, and in an array in the root switch. The \textit{``densification''} of sparse data depends on the number of non-overlapping indexes, and we analyze these effects at scale in Section~\ref{sec:evaluation:sparse}.

We can distribute the computation to the cores using the same solutions we described in Section~\ref{sec:parallelism} for the dense case: single buffer, multiple buffers, and tree aggregation. We can also use the same models, with a few caveats. First, for sparse allreduce, the term $P$ (i.e., the number of packets expected for a given block) represents the average case because, as described earlier, each block can be composed of more or less than $P$ packets. Moreover, storing $N$ elements might require more than $N$ memory cells, depending on the data structure used to store the partially aggregated data. 

\subsection{Evaluation}\label{sec:evaluation:sparse}
\begin{figure}[ht]
    \centering
    \includegraphics[width=\columnwidth]{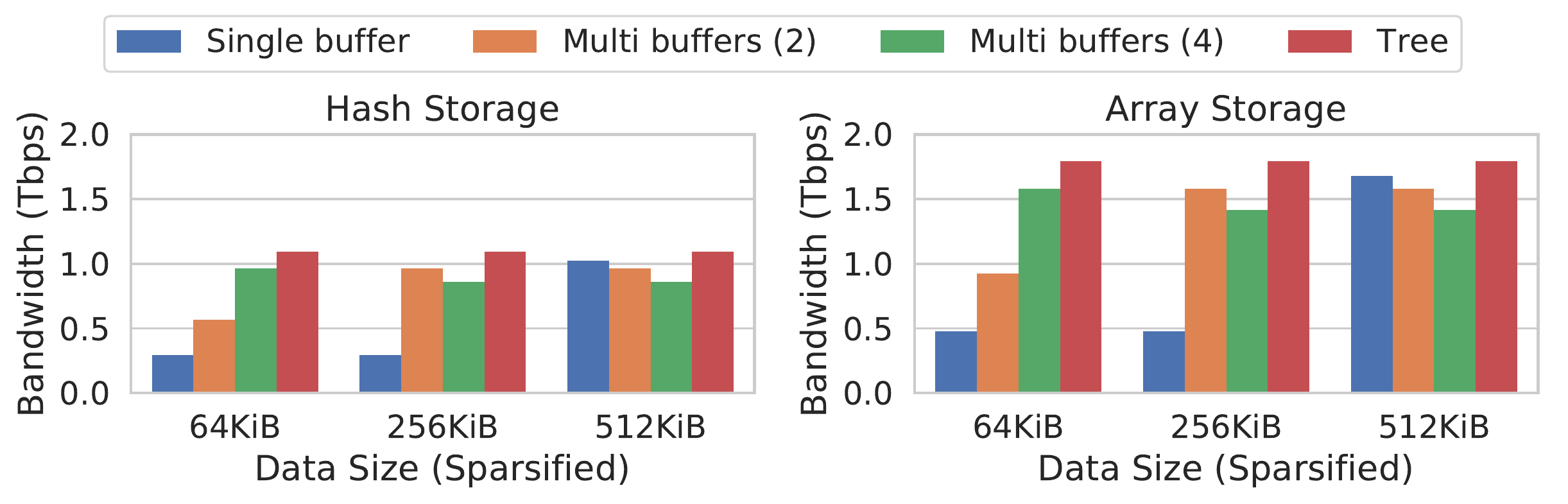}
    \caption{Modeled bandwidth for the \flare sparse allreduce.}
    \label{fig:modeled_sparse}
\end{figure}

First, we report the modeled bandwidth for both the hash and array storage in Figure~{\ref{fig:modeled_sparse}}, assuming a 10\% density. We observe a lower bandwidth for the sparse allreduce compared to the dense one, due to the more complex operations that need to be executed by the handler to store both the indexes and the data. However, as we will show later, because less data is transmitted, this still allows to obtain a performance advantage compared to both host-based and in-network allreduce.

\begin{figure}[ht]
    \centering
    \includegraphics[width=\columnwidth]{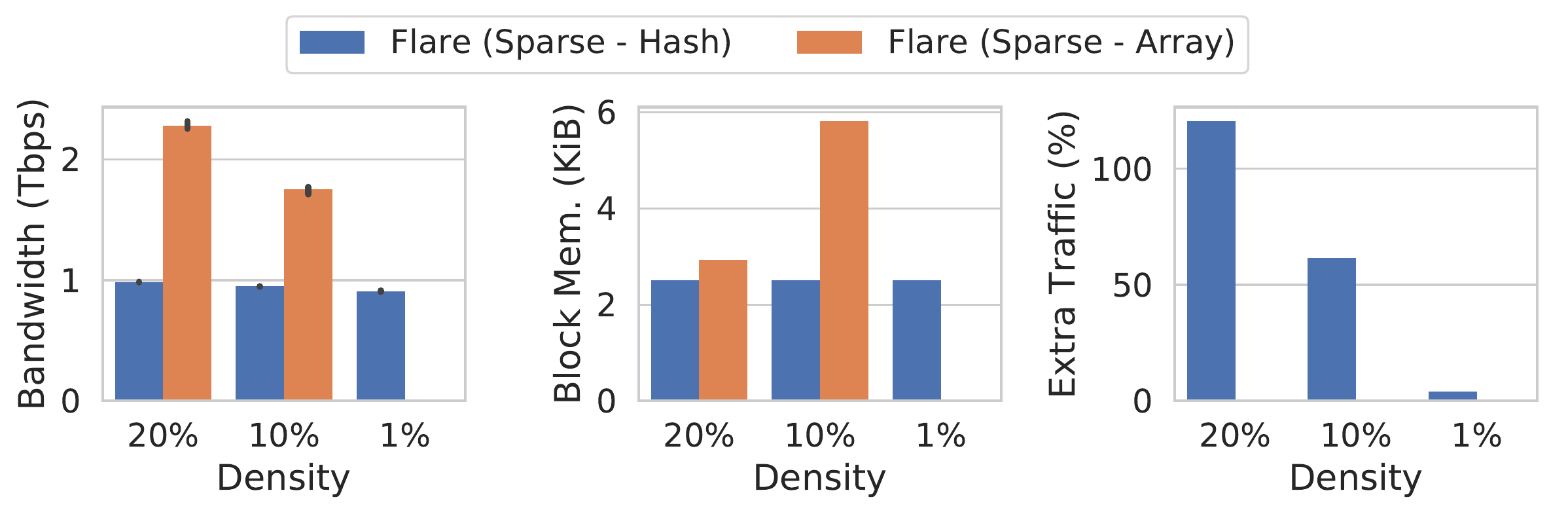}
    \caption{Simulated bandwidth, memory per block, and extra generated traffic for the \flare sparse data allreduce, for different density and storage types.}
    \label{fig:simulated_sparse}
\end{figure}

Then, we report the results for a 1MiB allreduce in Figure~\ref{fig:simulated_sparse}, analyzing the performance when using both the hash and the array storage, and for different data \textit{density}. Deep learning applications typically sparsify the data by only communicating the top 0.1\% or 1\% elements of the original data~\cite{sparcml}, whereas some graph processing algorithms might retain up to 20\% of the data~\cite{6957236,DBLP:journals/corr/ZhaoC13}. We observe that, as expected, storing the data as a contiguous array increases the bandwidth and the memory occupied by each block compared to using a hash table. We do not report the data about array storage with 1\% density because this requires a 600KiB array for each block (with, on average, only 6KiB non-zero elements) and all the concurrently processed blocks do not fit in \flare memory. Hash table storage is characterized by a constant bandwidth and memory occupancy independently from the density. Indeed, when data is stored in a hash table \flare always executes a number of instructions that only depend on the size of the packet. On the other hand, for lower density the array storage requires storing a larger array, thus increasing the number of cycles needed to flush it when the reduction is over.

However, to avoid costly collision management, with hash storage when there is a collision the element is put in a spill buffer, that when full is sent over the network. This leads to extra network traffic, because data is split between the hash table and the spill buffer, and the switch might not fully aggregate the data. We report the impact of spilling on the network traffic in Figure~\ref{fig:simulated_sparse}. For example, we observe that for $20\%$ data density, spilling doubles the network traffic. On the other hand, array storage never generates extra traffic, and packets are only sent when the block has been fully reduced.

To analyze the performance of the \flare sparse allreduce at scale, we extended the SST simulator~\cite{sst} so that the switch can modify in-transit packets, and we implemented both dense and sparse \flare in-network allreduce, and also the host-based ring allreduce. Moreover, we also implemented on top of SST the SparCML host-based sparse allreduce~\cite{sparcml}. We then tuned the simulator parameters so that the bandwidth of the switches matches with that obtained through the cycle-accurate PsPIN simulator. Both the PsPIN-based simulator and the extended SST simulator are provided in the artifact.

\begin{figure}[ht]
    \centering
    \includegraphics[width=\columnwidth]{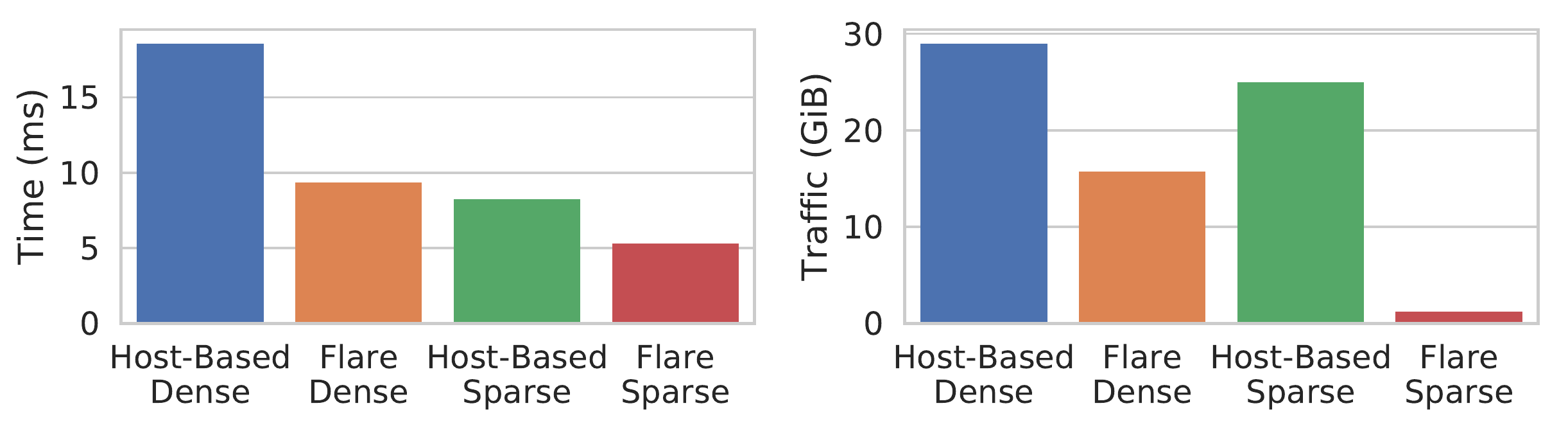}
    \caption{Execution time and network traffic for a 64 nodes allreduce executed on a simulated 2-levels fat tree network. The data to be reduced contains the gradients exchanged during a sparsified ResNet50 training iteration.}
    \label{fig:sparse_sst}
\end{figure}

As we discussed in Section~{\ref{sec:sparsity}}, \textsc{Flare} performance depends on the amount of overlapping indexes. To simulate a realistic scenario, we gathered the data exchanged during a Resnet50~{\cite{7780459}} training iteration executed on 64 nodes using SparCML, and we send the same data in SST. Each host works on a 100MiB vector of floating point values. For sparse allreduces, the data is split in buckets of 512 values, and one single value is sent for each bucket (${\sim} 0.2\%$ density). We reproduce the same data on a simulated 2-level fat tree network~{\cite{fattree}} built with 8-port 100Gbps switches, connecting 64 nodes with 100Gbps network interfaces. We report in Figure~{\ref{fig:sparse_sst}} the results of our analysis, by showing the time the hosts need to complete the allreduce and the total number of bytes that traversed the network.

First, we observe that the in-network dense allreduce leads to more than 2x speedup compared to the host-based dense allreduce, and to a 2x reduction in the network traffic. We also observe that the host-based sparse allreduce prodives slightly better performance than the in-network dense allreduce. However, it also generates more traffic because, even if less data is sent by the hosts, that data traverses more hops. Eventually, we observe that the in-network \textsc{Flare} sparse allreduce provides further advantages compared to both host-based sparse allreduce and in-network dense allreduce, both in terms of performance and network traffic. This leads to performance improvement up to $35\%$ and traffic reduction up 20x compared to the SparCML host-based sparse allreduce, and to performance improvement up to $43\%$ and traffic reduction up to 13x compared to the in-network \textsc{Flare} dense allreduce.

\section{Discussion}
\paragraph{Support for other collectives}
Although we considered in this work the allreduce collective operation, other collectives like \textit{reduce}, \textit{broadcast}, and \textit{barrier} can also be accelerated with \textsc{Flare}. For example, a \textit{barrier} can simply be implemented as an in-network \textit{allreduce} with 0-bytes data.
Moreover, in some frameworks like Horovod~\cite{horovod} each process might have multiple outstanding allreduce operations, and each rank might issue those operations in a different order, potentially leading to deadlock~\cite{kurth}. To address this problem, an additional synchronization step between the ranks is required, and this step could be offloaded to \textsc{Flare} as well, together with the allreduce operation.

\paragraph{Limitations}
As we shown in Section~\ref{sec:evaluation:sparse}, the bandwidth achieved by the in-network sparse allreduce is lower than its dense counterpart, due to the higher latency required for processing and aggregating the sparse data. In our experiments, the lower bandwidth was compensated by the reduction in the amount of processed data, and the \flare in-network sparse allreduce outperformed the SparCML host-based sparse allreduce. However, we believe that there is still space for improvement, either by optimizing the handlers code or by introducing hardware support to optimize indirect memory accesses~\cite{date-issr}.

\section{Conclusions}
In this work, we introduced \flare, an architecture for flexible in-network data reduction. \flare is based on the open-source PsPIN RISC-V architecture, and allows running custom packet processing handlers on network packets. \flare also includes a set of aggregation algorithms with different performance and memory occupancy tradeoffs. We modeled and analyzed in detail each of these algorithms, in terms of bandwidth and memory occupancy. We then implemented the aggregation algorithms in the PsPIN cycle-accurate simulator, and analyzed their performance for different allreduce sizes and datatypes, comparing them with state-of-the-art in-network aggregation approaches such as SHARP and SwitchML. Moreover, we also designed and implemented the first (to the best of our knowledge) in-network sparse allreduce algorithm. We implemented the algorithm in the PsPIN and SST simulators, showing performance improvements and network traffic reduction compared to in-network dense allreduce and host-based sparse allreduce.

\section*{Acknowledgments}
This work has been partially funded by the EPIGRAM-HS project under grant agreement no. 801039, and the European Project RED-SEA under grant agreement no.  955776. Daniele De Sensi is supported by an ETH Postdoctoral Fellowship (19-2 FEL-50).


\bibliographystyle{unsrt}
\bibliography{bib}

\begin{thebibliography}{10}

\bibitem{mpi}
Message~Passing Forum.
\newblock Mpi: A message-passing interface standard.
\newblock Technical report, USA, 1994.

\bibitem{mpi-characterization}
Sudheer Chunduri, Scott Parker, Pavan Balaji, Kevin Harms, and Kalyan Kumaran.
\newblock Characterization of mpi usage on a production supercomputer.
\newblock In {\em Proceedings of the International Conference for High
  Performance Computing, Networking, Storage, and Analysis}, SC '18. IEEE
  Press, 2018.

\bibitem{milc}
Steven Gottlieb, W.~Liu, {William D} Toussaint, {R. L.} Renken, and {R. L.}
  Sugar.
\newblock Hybrid-molecular-dynamics algorithms for the numerical simulation of
  quantum chromodynamics.
\newblock {\em Physical review D: Particles and fields}, 35(8):2531--2542,
  1987.

\bibitem{10.1145/3320060}
Tal Ben-Nun and Torsten Hoefler.
\newblock Demystifying parallel and distributed deep learning: An in-depth
  concurrency analysis.
\newblock {\em ACM Comput. Surv.}, 52(4), August 2019.

\bibitem{6957236}
H.~{Zhao} and J.~{Canny}.
\newblock Kylix: A sparse allreduce for commodity clusters.
\newblock In {\em 2014 43rd International Conference on Parallel Processing},
  pages 273--282, 2014.

\bibitem{DBLP:journals/corr/ZhaoC13}
Huasha Zhao and John~F. Canny.
\newblock Sparse allreduce: Efficient scalable communication for power-law
  data.
\newblock {\em CoRR}, abs/1312.3020, 2013.

\bibitem{10.1007/978-3-642-03770-2_30}
Torsten Hoefler, Andrew Lumsdaine, and Jack Dongarra.
\newblock Towards efficient mapreduce using mpi.
\newblock In Matti Ropo, Jan Westerholm, and Jack Dongarra, editors, {\em
  Recent Advances in Parallel Virtual Machine and Message Passing Interface},
  pages 240--249, Berlin, Heidelberg, 2009. Springer Berlin Heidelberg.

\bibitem{bwoptimalallreduce}
Pitch Patarasuk and Xin Yuan.
\newblock Bandwidth optimal all-reduce algorithms for clusters of workstations.
\newblock {\em Journal of Parallel and Distributed Computing}, 69(2):117 --
  124, 2009.

\bibitem{sharp}
Richard~L. Graham, Devendar Bureddy, Pak Lui, Hal Rosenstock, Gilad Shainer,
  Gil Bloch, Dror Goldenerg, Mike Dubman, Sasha Kotchubievsky, Vladimir
  Koushnir, Lion Levi, Alex Margolin, Tamir Ronen, Alexander Shpiner, Oded
  Wertheim, and Eitan Zahavi.
\newblock {Scalable Hierarchical Aggregation Protocol (SHArP): A Hardware
  Architecture for Efficient Data Reduction}.
\newblock In {\em Proceedings of COM-HPC 2016: 1st Workshop on Optimization of
  Communication in HPC Runtime Systems - Held in conjunction with SC 2016: The
  International Conference for High Performance Computing, Networking, Storage
  and Analysis}, pages 1--10. Institute of Electrical and Electronics Engineers
  Inc., jan 2017.

\bibitem{nvidiaar}
Benjamin Klenk, Nan Jiang, Greg Thorson, and Larry Dennison.
\newblock An in-network architecture for accelerating shared-memory
  multiprocessor collectives.
\newblock In {\em Proceedings of the ACM/IEEE 47th Annual International
  Symposium on Computer Architecture}, ISCA '20, page 996–1009. IEEE Press,
  2020.

\bibitem{switchml}
Amedeo Sapio, Marco Canini, Chen-Yu Ho, Jacob Nelson, Panos Kalnis, Changhoon
  Kim, Arvind Krishnamurthy, Masoud Moshref, Dan R.~K. Ports, and Peter
  Richtárik.
\newblock Scaling distributed machine learning with in-network aggregation.
\newblock In {\em 18th {USENIX} Symposium on Networked Systems Design and
  Implementation ({NSDI} 21)}. {USENIX} Association, April 2021.

\bibitem{10.1145/1816038.1816004}
Dennis Abts, Michael~R. Marty, Philip~M. Wells, Peter Klausler, and Hong Liu.
\newblock Energy proportional datacenter networks.
\newblock {\em SIGARCH Comput. Archit. News}, 38(3):338–347, June 2010.

\bibitem{pspin}
Salvatore Di~Girolamo, Andreas Kurth, Alexandru Calotoiu, Thomas Benz, Timo
  Schneider, Jakub Beranek, Luca Benini, and Torsten Hoefler.
\newblock A risc-v in-network accelerator for flexible high-performance
  low-power packet processing.
\newblock In {\em 2021 ACM/IEEE 48th Annual International Symposium on Computer
  Architecture (ISCA)}, 2021.

\bibitem{Waterman:EECS-2014-54}
Andrew Waterman, Yunsup Lee, David~A. Patterson, and Krste Asanović.
\newblock The risc-v instruction set manual, volume i: User-level isa, version
  2.0.
\newblock Technical Report UCB/EECS-2014-54, EECS Department, University of
  California, Berkeley, May 2014.

\bibitem{spin}
Torsten Hoefler, Salvatore Di~Girolamo, Konstantin Taranov, Ryan~E. Grant, and
  Ron Brightwell.
\newblock Spin: High-performance streaming processing in the network.
\newblock In {\em Proceedings of the International Conference for High
  Performance Computing, Networking, Storage and Analysis}, SC '17, New York,
  NY, USA, 2017. Association for Computing Machinery.

\bibitem{sharp2}
Richard~L. Graham, Lion Levi, Devendar Burredy, Gil Bloch, Gilad Shainer, David
  Cho, George Elias, Daniel Klein, Joshua Ladd, Ophir Maor, Ami Marelli,
  Valentin Petrov, Evyatar Romlet, Yong Qin, and Ido Zemah.
\newblock {Scalable Hierarchical Aggregation and Reduction Protocol (SHARP)TM
  Streaming-Aggregation Hardware Design and Evaluation}.
\newblock In {\em Lecture Notes in Computer Science (including subseries
  Lecture Notes in Artificial Intelligence and Lecture Notes in
  Bioinformatics)}, volume 12151 LNCS, pages 41--59. Springer, jun 2020.

\bibitem{crayxcpdf}
Bob Alverson, Edwin Froese, Larry Kaplan, and Duncan Roweth.
\newblock Cray xc series network.
\newblock {\em Cray Inc., White Paper WP-Aries01-1112}, 2012.

\bibitem{tofu}
Y.~{Ajima}, T.~{Kawashima}, T.~{Okamoto}, N.~{Shida}, K.~{Hirai}, T.~{Shimizu},
  S.~{Hiramoto}, Y.~{Ikeda}, T.~{Yoshikawa}, K.~{Uchida}, and T.~{Inoue}.
\newblock The tofu interconnect d.
\newblock In {\em 2018 IEEE International Conference on Cluster Computing
  (CLUSTER)}, pages 646--654, 2018.

\bibitem{percs}
B.~{Arimilli}, R.~{Arimilli}, V.~{Chung}, S.~{Clark}, W.~{Denzel}, B.~{Drerup},
  T.~{Hoefler}, J.~{Joyner}, J.~{Lewis}, J.~{Li}, N.~{Ni}, and R.~{Rajamony}.
\newblock The percs high-performance interconnect.
\newblock In {\em 2010 18th IEEE Symposium on High Performance Interconnects},
  pages 75--82, 2010.

\bibitem{percspatent}
B.~{Arimilli}, Bernard~C. Drerup, Paul~F. Lecocq, and Hanhong Xue.
\newblock Collective acceleration unit tree structure, U.S. Patent US8756270B2,
  17/06/2014.

\bibitem{anton}
J.~P. {Grossman}, B.~{Towles}, B.~{Greskamp}, and D.~E. {Shaw}.
\newblock Filtering, reductions and synchronization in the anton 2 network.
\newblock In {\em 2015 IEEE International Parallel and Distributed Processing
  Symposium}, pages 860--870, 2015.

\bibitem{panama}
Nadeen Gebara, Paolo Costa, and Manya Ghobadi.
\newblock Panama: In-network aggregation for shared machine learning clusters.
\newblock In {\em Conference on Machine Learning and Systems (MLSys)}, April
  2021.

\bibitem{netreduce}
Shuo Liu, Qiaoling Wang, Junyi Zhang, Qinliang Lin, Yao Liu, Meng Xu, Ray C.~C.
  Chueng, and Jianfei He.
\newblock Netreduce: Rdma-compatible in-network reduction for distributed dnn
  training acceleration, 2020.

\bibitem{atp}
ChonLam Lao, Yanfang Le, Kshiteej Mahajan, Yixi Chen, Wenfei Wu, Aditya Akella,
  and Michael Swift.
\newblock {ATP}: In-network aggregation for multi-tenant learning.
\newblock In {\em 18th {USENIX} Symposium on Networked Systems Design and
  Implementation ({NSDI} 21)}. {USENIX} Association, April 2021.

\bibitem{omnireduce}
Jiawei Fei, Chen-Yu Ho, Atal~Narayan Sahu, Marco Canini, and Amedeo Sapio.
\newblock {Efficient Sparse Collective Communication and its application to
  Accelerate Distributed Deep Learning}.
\newblock Technical report, KAUST, Sep 2020.

\bibitem{rmt}
Pat Bosshart, Glen Gibb, Hun-Seok Kim, George Varghese, Nick McKeown, Martin
  Izzard, Fernando Mujica, and Mark Horowitz.
\newblock Forwarding metamorphosis: Fast programmable match-action processing
  in hardware for sdn.
\newblock {\em SIGCOMM Comput. Commun. Rev.}, 43(4):99–110, August 2013.

\bibitem{tofino}
Intel.
\newblock {Intel Tofino Series}.
\newblock
  \url{https://www.intel.com/content/www/us/en/products/network-io/programmable-ethernet-switch.html},
  mar 2021.

\bibitem{intelfm}
Recep~Ozdag Intel.
\newblock {Intel (R) Ethernet Switch FM6000 Series -- Software Defined
  Networking}.
\newblock
  https://people.ucsc.edu/~warner/Bufs/ethernet-switch-fm6000-sdn-paper.pdf,
  mar 2021.

\bibitem{8850761}
R.~{Bifulco} and G.~{Rétvári}.
\newblock A survey on the programmable data plane: Abstractions, architectures,
  and open problems.
\newblock In {\em 2018 IEEE 19th International Conference on High Performance
  Switching and Routing (HPSR)}, pages 1--7, 2018.

\bibitem{p4}
Pat Bosshart, Dan Daly, Glen Gibb, Martin Izzard, Nick McKeown, Jennifer
  Rexford, Cole Schlesinger, Dan Talayco, Amin Vahdat, George Varghese, and
  David Walker.
\newblock P4: Programming protocol-independent packet processors.
\newblock {\em SIGCOMM Comput. Commun. Rev.}, 44(3):87–95, July 2014.

\bibitem{survey-p4-1}
Frederik Hauser, Marco Häberle, Daniel Merling, Steffen Lindner, Vladimir
  Gurevich, Florian Zeiger, Reinhard Frank, and Michael Menth.
\newblock A survey on data plane programming with p4: Fundamentals, advances,
  and applied research, 2021.

\bibitem{survey-p4-2}
Elie~F. Kfoury, Jorge Crichigno, and Elias Bou-Harb.
\newblock An exhaustive survey on p4 programmable data plane switches:
  Taxonomy, applications, challenges, and future trends, 2021.

\bibitem{10.1145/3317550.3321439}
Dan R.~K. Ports and Jacob Nelson.
\newblock When should the network be the computer?
\newblock In {\em Proceedings of the Workshop on Hot Topics in Operating
  Systems}, HotOS '19, page 209–215, New York, NY, USA, 2019. Association for
  Computing Machinery.

\bibitem{201474}
Naveen~Kr. Sharma, Antoine Kaufmann, Thomas Anderson, Arvind Krishnamurthy,
  Jacob Nelson, and Simon Peter.
\newblock Evaluating the power of flexible packet processing for network
  resource allocation.
\newblock In {\em 14th {USENIX} Symposium on Networked Systems Design and
  Implementation ({NSDI} 17)}, pages 67--82, Boston, MA, March 2017. {USENIX}
  Association.

\bibitem{hoefler2021sparsity}
Torsten Hoefler, Dan Alistarh, Tal Ben-Nun, Nikoli Dryden, and Alexandra Peste.
\newblock Sparsity in deep learning: Pruning and growth for efficient inference
  and training in neural networks, 2021.

\bibitem{quantization}
Song Han, Huizi Mao, and William~J. Dally.
\newblock Deep compression: Compressing deep neural networks with pruning,
  trained quantization and huffman coding, 2016.

\bibitem{10.5555/3018874.3018875}
Nikoli Dryden, Sam~Ade Jacobs, Tim Moon, and Brian Van~Essen.
\newblock Communication quantization for data-parallel training of deep neural
  networks.
\newblock In {\em Proceedings of the Workshop on Machine Learning in High
  Performance Computing Environments}, MLHPC '16, page 1–8. IEEE Press, 2016.

\bibitem{37631}
Vincent Vanhoucke, Andrew Senior, and Mark~Z. Mao.
\newblock Improving the speed of neural networks on cpus.
\newblock In {\em Deep Learning and Unsupervised Feature Learning Workshop,
  NIPS 2011}, 2011.

\bibitem{sparcml}
Cedric Renggli, Saleh Ashkboos, Mehdi Aghagolzadeh, Dan Alistarh, and Torsten
  Hoefler.
\newblock Sparcml: High-performance sparse communication for machine learning.
\newblock In {\em Proceedings of the International Conference for High
  Performance Computing, Networking, Storage and Analysis}, SC '19, New York,
  NY, USA, 2019. Association for Computing Machinery.

\bibitem{bams-htor}
Christoph Schär, Oliver Fuhrer, Andrea Arteaga, Nikolina Ban, Christophe
  Charpilloz, Salvatore~Di Girolamo, Laureline Hentgen, Torsten Hoefler, Xavier
  Lapillonne, David Leutwyler, Katherine Osterried, Davide Panosetti, Stefan
  Rüdisühli, Linda Schlemmer, Thomas Schulthess, Michael Sprenger, Stefano
  Ubbiali, and Heini Wernli.
\newblock {Kilometer-scale climate models: Prospects and challenges}.
\newblock {\em Bulletin of the American Meteorological Society}, 100(12), Dec.
  2019.
\newblock Early Online Release.

\bibitem{weather}
Edward~N. Lorenz.
\newblock The predictability of a flow which possesses many scales of motion.
\newblock {\em Tellus}, 21(3):289--307, 1969.

\bibitem{Geyer2021}
Beate Geyer, Thomas Ludwig, and Hans von Storch.
\newblock Limits of reproducibility and hydrodynamic noise in atmospheric
  regional modelling.
\newblock {\em Communications Earth {\&} Environment}, 2(1):17, Jan 2021.

\bibitem{COLLANGE201583}
Sylvain Collange, David Defour, Stef Graillat, and Roman Iakymchuk.
\newblock Numerical reproducibility for the parallel reduction on multi- and
  many-core architectures.
\newblock {\em Parallel Computing}, 49:83--97, 2015.

\bibitem{6831947}
P.~{Balaji} and D.~{Kimpe}.
\newblock On the reproducibility of mpi reduction operations.
\newblock In {\em 2013 IEEE 10th International Conference on High Performance
  Computing and Communications 2013 IEEE International Conference on Embedded
  and Ubiquitous Computing}, pages 407--414, 2013.

\bibitem{6812157}
N.~Revol and P.~Theveny.
\newblock Numerical reproducibility and parallel computations: Issues for
  interval algorithms.
\newblock {\em IEEE Transactions on Computers}, 63(08):1915--1924, aug 2014.

\bibitem{Villa_effectsof}
Oreste Villa, Vidhya Gurumoorthi, and Sriram Krishnamoorthy.
\newblock Effects of floating-point nonassociativity on numerical computations
  on massively multithreaded systems.
\newblock In {\em In CUG 2009 Proceedings}, pages 1--11, 2009.

\bibitem{ZOLFAGHARI2020102910}
Hesam Zolfaghari, Davide Rossi, and Jari Nurmi.
\newblock A custom processor for protocol-independent packet parsing.
\newblock {\em Microprocessors and Microsystems}, 72:102910, 2020.

\bibitem{6665172}
G.~{Gibb}, G.~{Varghese}, M.~{Horowitz}, and N.~{McKeown}.
\newblock Design principles for packet parsers.
\newblock In {\em Architectures for Networking and Communications Systems},
  pages 13--24, 2013.

\bibitem{sdn}
Nick Feamster, Jennifer Rexford, and Ellen Zegura.
\newblock The road to sdn: An intellectual history of programmable networks.
\newblock {\em SIGCOMM Comput. Commun. Rev.}, 44(2):87–98, April 2014.

\bibitem{rfc894}
{Internet Control Message Protocol}.
\newblock RFC 894, April 1984.

\bibitem{domino}
Anirudh Sivaraman, Alvin Cheung, Mihai Budiu, Changhoon Kim, Mohammad Alizadeh,
  Hari Balakrishnan, George Varghese, Nick McKeown, and Steve Licking.
\newblock Packet transactions: High-level programming for line-rate switches.
\newblock In {\em Proceedings of the 2016 ACM SIGCOMM Conference}, SIGCOMM '16,
  page 15–28, New York, NY, USA, 2016. Association for Computing Machinery.

\bibitem{flowblaze}
Salvatore Pontarelli, Roberto Bifulco, Marco Bonola, Carmelo Cascone, Marco
  Spaziani, Valerio Bruschi, Davide Sanvito, Giuseppe Siracusano, Antonio
  Capone, Michio Honda, Felipe Huici, and Giuseppe Siracusano.
\newblock Flowblaze: Stateful packet processing in hardware.
\newblock In {\em 16th {USENIX} Symposium on Networked Systems Design and
  Implementation ({NSDI} 19)}, pages 531--548, Boston, MA, February 2019.
  {USENIX} Association.

\bibitem{pulp}
Davide Rossi, Francesco Conti, Andrea Marongiu, Antonio Pullini, Igor Loi,
  Michael Gautschi, Giuseppe Tagliavini, Alessandro Capotondi, Philippe
  Flatresse, and Luca Benini.
\newblock {PULP: A parallel ultra low power platform for next generation IoT
  applications}.
\newblock In {\em 2015 IEEE Hot Chips 27 Symposium (HCS)}, pages 1--39. IEEE,
  2015.

\bibitem{7864441}
M.~{Gautschi}, P.~D. {Schiavone}, A.~{Traber}, I.~{Loi}, A.~{Pullini},
  D.~{Rossi}, E.~{Flamand}, F.~K. {Gürkaynak}, and L.~{Benini}.
\newblock Near-threshold risc-v core with dsp extensions for scalable iot
  endpoint devices.
\newblock {\em IEEE Transactions on Very Large Scale Integration (VLSI)
  Systems}, 25(10):2700--2713, 2017.

\bibitem{fpnew}
S.~Mach, F.~Schuiki, F.~Zaruba, and L.~Benini.
\newblock Fpnew: An open-source multiformat floating-point unit architecture
  for energy-proportional transprecision computing.
\newblock {\em IEEE Transactions on Very Large Scale Integration (VLSI)
  Systems}, 29(04):774--787, apr 2021.

\bibitem{arealong}
H.~{Zolfaghari}, D.~{Rossi}, W.~{Cerroni}, H.~{Okuhara}, C.~{Raffaelli}, and
  J.~{Nurmi}.
\newblock Flexible software-defined packet processing using low-area hardware.
\newblock {\em IEEE Access}, 8:98929--98945, 2020.

\bibitem{slingshot}
Daniele De~Sensi, Salvatore Di~Girolamo, Kim~H. McMahon, Duncan Roweth, and
  Torsten Hoefler.
\newblock An in-depth analysis of the slingshot interconnect.
\newblock In {\em Proceedings of the International Conference for High
  Performance Computing, Networking, Storage and Analysis}, SC '20. IEEE Press,
  2020.

\bibitem{tomahawk4}
Broadcom.
\newblock {Tomahawk4 BCM56990 Series}.
\newblock
  https://www.broadcom.com/products/ethernet-connectivity/switching/strataxgs/bcm56990-series,
  mar 2019.

\bibitem{littleslaw}
John D.~C. Little.
\newblock {A Proof for the Queuing Formula: $L = \lambda W$}.
\newblock {\em Oper. Res.}, 9(3):383–387, June 1961.

\bibitem{li2020taming}
Shigang Li, Tal Ben-Nun, Salvatore~Di Girolamo, Dan Alistarh, and Torsten
  Hoefler.
\newblock Taming unbalanced training workloads in deep learning with partial
  collective operations.
\newblock In {\em Proceedings of the 25th ACM SIGPLAN Symposium on Principles
  and Practice of Parallel Programming}, pages 45--61, 2020.

\bibitem{imbalance}
Ahmad Faraj, Pitch Patarasuk, and Xin Yuan.
\newblock A study of process arrival patterns for mpi collective operations.
\newblock {\em Int. J. Parallel Program.}, 36(6):543--570, December 2008.

\bibitem{10.1145/3295500.3356196}
Daniele De~Sensi, Salvatore Di~Girolamo, and Torsten Hoefler.
\newblock Mitigating network noise on dragonfly networks through
  application-aware routing.
\newblock In {\em Proceedings of the International Conference for High
  Performance Computing, Networking, Storage and Analysis}, SC '19, New York,
  NY, USA, 2019. Association for Computing Machinery.

\bibitem{allreduce}
T.~Groves, Y.~Gu, and N.~J. Wright.
\newblock Understanding performance variability on the aries dragonfly network.
\newblock In {\em 2017 IEEE International Conference on Cluster Computing
  (CLUSTER)}, pages 809--813, Sept 2017.

\bibitem{htornnoise}
T.~Hoefler, T.~Schneider, and A.~Lumsdaine.
\newblock The impact of network noise at large-scale communication performance.
\newblock In {\em 2009 IEEE International Symposium on Parallel Distributed
  Processing}, pages 1--8, May 2009.

\bibitem{Chunduri:2017:RVX:3126908.3126926}
Sudheer Chunduri, Kevin Harms, Scott Parker, Vitali Morozov, Samuel Oshin,
  Naveen Cherukuri, and Kalyan Kumaran.
\newblock Run-to-run variability on xeon phi based cray xc systems.
\newblock In {\em Proceedings of the International Conference for High
  Performance Computing, Networking, Storage and Analysis}, SC '17, pages
  52:1--52:13, New York, NY, USA, 2017. ACM.

\bibitem{1526010}
D.~{Skinner} and W.~{Kramer}.
\newblock Understanding the causes of performance variability in hpc workloads.
\newblock In {\em IEEE International. 2005 Proceedings of the IEEE Workload
  Characterization Symposium, 2005.}, pages 137--149, Oct 2005.

\bibitem{osnoise}
Torsten Hoefler, Timo Schneider, and Andrew Lumsdaine.
\newblock Characterizing the influence of system noise on large-scale
  applications by simulation.
\newblock In {\em Proceedings of the 2010 ACM/IEEE International Conference for
  High Performance Computing, Networking, Storage and Analysis}, SC '10, pages
  1--11, Washington, DC, USA, 2010. IEEE Computer Society.

\bibitem{6012894}
A.~{Morari}, R.~{Gioiosa}, R.~W. {Wisniewski}, F.~J. {Cazorla}, and
  M.~{Valero}.
\newblock A quantitative analysis of os noise.
\newblock In {\em 2011 IEEE International Parallel Distributed Processing
  Symposium}, pages 852--863, 2011.

\bibitem{ibm:tm}
Bogdan Prisacari, German Rodriguez, Philip Heidelberger, Dong Chen, Cyriel
  Minkenberg, and Torsten Hoefler.
\newblock Efficient task placement and routing of nearest neighbor exchanges in
  dragonfly networks.
\newblock In {\em Proceedings of the 23rd International Symposium on
  High-performance Parallel and Distributed Computing}, HPDC '14, pages
  129--140, New York, NY, USA, 2014. ACM.

\bibitem{fattree:sc18}
Samuel~D. Pollard, Nikhil Jain, Stephen Herbein, and Abhinav Bhatele.
\newblock Evaluation of an interference-free node allocation policy on fat-tree
  clusters.
\newblock In {\em Proceedings of the International Conference for High
  Performance Computing, Networking, Storage, and Analysis}, SC '18, pages
  26:1--26:13, Piscataway, NJ, USA, 2018. IEEE Press.

\bibitem{doi:10.1142/S0129626409000419}
Abhinav Bhatele and Laxmikant~V. Kal\'{e}.
\newblock Quantifying network contention on large parallel machines.
\newblock {\em Parallel Processing Letters}, 19(04):553--572, 2009.

\bibitem{bully}
X.~Yang, J.~Jenkins, M.~Mubarak, R.~B. Ross, and Z.~Lan.
\newblock Watch out for the bully! job interference study on dragonfly network.
\newblock In {\em SC '16: Proceedings of the International Conference for High
  Performance Computing, Networking, Storage and Analysis}, pages 750--760, Nov
  2016.

\bibitem{8425264}
X.~{Wang}, M.~{Mubarak}, X.~{Yang}, R.~B. {Ross}, and Z.~{Lan}.
\newblock Trade-off study of localizing communication and balancing network
  traffic on a dragonfly system.
\newblock In {\em 2018 IEEE International Parallel and Distributed Processing
  Symposium (IPDPS)}, pages 1113--1122, May 2018.

\bibitem{sc18}
Staci~A. Smith, Clara~E. Cromey, David~K. Lowenthal, Jens Domke, Nikhil Jain,
  Jayaraman~J. Thiagarajan, and Abhinav Bhatele.
\newblock Mitigating inter-job interference using adaptive flow-aware routing.
\newblock In {\em Proceedings of the International Conference for High
  Performance Computing, Networking, Storage and Analysis}, SC '18, 2018.

\bibitem{atun}
A.~{Kurth}, S.~{Riedel}, F.~{Zaruba}, T.~{Hoefler}, and L.~{Benini}.
\newblock Atuns: Modular and scalable support for atomic operations in a shared
  memory multiprocessor.
\newblock In {\em 2020 57th ACM/IEEE Design Automation Conference (DAC)}, pages
  1--6, 2020.

\bibitem{atomics}
H.~{Schweizer}, M.~{Besta}, and T.~{Hoefler}.
\newblock Evaluating the cost of atomic operations on modern architectures.
\newblock In {\em 2015 International Conference on Parallel Architecture and
  Compilation (PACT)}, pages 445--456, 2015.

\bibitem{fireforget}
Maciej Besta and Torsten Hoefler.
\newblock Accelerating irregular computations with hardware transactional
  memory and active messages.
\newblock In {\em Proceedings of the 24th International Symposium on
  High-Performance Parallel and Distributed Computing}, HPDC '15, page
  161–172, New York, NY, USA, 2015. Association for Computing Machinery.

\bibitem{quantum}
Mellanox.
\newblock {Mellanox Quantum Switches}.
\newblock
  \url{https://www.mellanox.com/products/infiniband-switches-ic/quantum}, mar
  2019.

\bibitem{sst}
A.~F. Rodrigues, K.~S. Hemmert, B.~W. Barrett, C.~Kersey, R.~Oldfield,
  M.~Weston, R.~Risen, J.~Cook, P.~Rosenfeld, E.~Cooper-Balis, and B.~Jacob.
\newblock The structural simulation toolkit.
\newblock {\em SIGMETRICS Perform. Eval. Rev.}, 38(4):37–42, March 2011.

\bibitem{7780459}
Kaiming He, Xiangyu Zhang, Shaoqing Ren, and Jian Sun.
\newblock Deep residual learning for image recognition.
\newblock In {\em 2016 IEEE Conference on Computer Vision and Pattern
  Recognition (CVPR)}, pages 770--778, 2016.

\bibitem{fattree}
C.~E. {Leiserson}.
\newblock Fat-trees: Universal networks for hardware-efficient supercomputing.
\newblock {\em IEEE Transactions on Computers}, C-34(10):892--901, 1985.

\bibitem{horovod}
Alexander Sergeev and Mike~Del Balso.
\newblock Horovod: fast and easy distributed deep learning in tensorflow, 2018.

\bibitem{kurth}
Thorsten Kurth, Sean Treichler, Joshua Romero, Mayur Mudigonda, Nathan Luehr,
  Everett Phillips, Ankur Mahesh, Michael Matheson, Jack Deslippe, Massimiliano
  Fatica, Prabhat, and Michael Houston.
\newblock Exascale deep learning for climate analytics.
\newblock In {\em Proceedings of the International Conference for High
  Performance Computing, Networking, Storage, and Analysis}, SC '18. IEEE
  Press, 2018.

\bibitem{date-issr}
Paul Scheffler, Florian Zaruba, Fabian Schuiki, Torsten Hoefler, and Luca
  Benini.
\newblock {Indirection Stream Semantic Register Architecture for Efficient
  Sparse-Dense Linear Algebra}.
\newblock 2021.

\end{thebibliography}

\end{document}